%% file: BPH-22-001_temp.tex
\pdfoutput=1
\documentclass[11pt,twoside,a4paper,cmspaper,final,collab]{cms-tdr}
\def\svnVersion{92a2878}\def\svnDate{2024/11/15}\def\cmsCernNoTag{CERN-EP-2024-172}\def\cmsCernDate{\today}\def\cmsMessage{Published in the Journal of High Energy Physics as \href{http://dx.doi.org/10.1007/JHEP10(2024)247}{\doi{10.1007/JHEP10(2024)247}.}}
\begin{document}\cmsNoteHeader{BPH-22-001}

\newlength\cmsTabSkip\setlength{\cmsTabSkip}{1ex}
\newcommand{\bd}        {\ensuremath{\PBz}\xspace}
\newcommand{\bs}        {\ensuremath{\PBzs}\xspace}
\newcommand{\bc}        {\ensuremath{\PBpc}\xspace}
\newcommand{\bsb}       {\PABzs}
\newcommand{\bdb}       {\PABz}
\newcommand{\chisq}     {\ensuremath{\chi^2}\xspace}
\newcommand{\ks}        {\ensuremath{\PKzS}\xspace}
\newcommand{\bsjpsiks}  {\ensuremath{\PBzs \to \PJGy\PKzS}\xspace}
\newcommand{\bdjpsiks}  {\ensuremath{\PBz \to \PJGy\PKzS}\xspace}
\newcommand{\jpsiks}    {\ensuremath{\PJGy\PKzS}\xspace}
\newcommand{\jpsi}      {\ensuremath{\PJGy}\xspace}
\newcommand{\st}        {\ensuremath{\sigma_t}\xspace}
\newcommand{\ps}        {\unit{ps}}
\newcommand{\mumu}      {\ensuremath{\PGmp\PGmm}\xspace}
\newcommand{\pipi}      {\ensuremath{\PGpp\PGpm}\xspace}
\newcommand{\ltoprpi}   {\ensuremath{\PgL \to \Pp \Pgpm}\xspace}
\DeclareRobustCommand{\Kstar}{\HepParticleResonanceFull{\PK}{}{\ast{}}{892}{}{0}\xspace}
\newcommand{\bmm}       {\ensuremath{\PBzs \to \MM}\xspace}
\newcommand{\bdmm}      {\ensuremath{\PBz \to \MM}\xspace}
\newcommand{\cp}        {\ensuremath{C\hspace{-.08em}P}\xspace}

\cmsNoteHeader{BPH-22-001}
\title{Measurement of the \bsjpsiks effective lifetime from proton-proton collisions at \texorpdfstring{$\sqrt{s} = 13\TeV$}{sqrt(s) = 13 TeV}}

\date{\today}

\abstract{The effective lifetime of the \bs meson in the decay \bsjpsiks is measured using data collected during 2016--2018 with the CMS detector in $\sqrt{s}=13\TeV$ proton-proton collisions at the LHC, corresponding to an integrated luminosity of 140\fbinv. The effective lifetime is determined by performing a two-dimensional unbinned maximum likelihood fit to the \bs meson invariant mass and proper decay time distributions. The resulting value of $1.59 \pm 0.07 \stat\pm 0.03 \syst \ps$ is the most precise measurement to date and is in good agreement with the expected value.}

\hypersetup{
pdfauthor={CMS Collaboration},
pdftitle={Measurement of the B0s to J/psi K0S effective lifetime from proton-proton collisions at sqrt(s) = 13 TeV},
pdfsubject={CMS},
pdfkeywords={CMS, lifetime, experimental results, flavor physics, CP violation}}

\maketitle 

\section{Introduction}
Measurements using \PQb hadron decays have dramatically improved our knowledge of the standard 
model (SM) flavor sector. An important aspect of \PB meson studies is determining whether
the SM correctly describes the violation of charge-parity (\cp) symmetry observed in 
nature. The oscillations of \bd and \bs mesons provide a valuable tool in the study of \cp violation~\cite{Albrecht:2024oyn}.
At hadron colliders, neutral \PB mesons are produced as flavor eigenstates, but they propagate as mass eigenstates,
referred to as the light and heavy states, which are linear combinations of the two flavor
states.
In the limit of \cp conservation in the flavor mixing, the mass and \cp eigenstates are identical.
These mass eigenstates have different lifetimes, both of which differ from the average lifetime of the corresponding \PB meson.

In the SM, quark flavor transitions occur through the weak interaction according to the Cabibbo-Kobayashi-Maskawa (CKM) matrix.
Unitarity triangles can be derived from this matrix, with a nonzero area indicating \cp violation~\cite{Cabibbo:1963yz, Kobayashi:1973fv}. Measurements of the angles 
and sides of the unitarity triangles can be used to constrain SM parameters and search for physics beyond the SM. The parameter 
$\beta$ is one of the angles of the most commonly used unitarity triangle and is most precisely determined by measuring the phase $\phi_\PQd$ 
in a time-dependent \cp asymmetry analysis of \bdjpsiks decays, where $\phi_\PQd$ arises from the interference between the direct decay of \bd and 
the decay after oscillating to \bdb. In the SM, $\phi_\PQd=2\beta$. However, the value of the phase measured in \bdjpsiks may not exactly 
equal $\phi_\PQd$ since there is a small contribution to the decay from doubly-Cabibbo-suppressed penguin decays, with a different phase~\cite{Ciuchini:2005mg,
Faller:2008zc, Jung:2012mp, Frings:2015eva}. Since the value of $\sin(2\beta)$ is now known to better than 2\%~\cite{LHCb:2023zcp}, understanding the penguin contribution is becoming more important in determining $\phi_\PQd$. 
This can be accomplished by studying the decay \bsjpsiks, where the penguin contribution is much larger, since it is not Cabibbo 
suppressed relative to the tree diagram, and it is related to the \bdjpsiks decay through the U-spin flavor symmetry of the 
strong interaction (swapping \PQs and \PQd quarks)~\cite{Fleischer:1999sj,DeBruyn:2010hh,DeBruyn:2014oga,Barel:2020jvf,Barel:2022wfr}.

\begin{figure}[htb]
  \centering
  \includegraphics[width=0.42\textwidth]{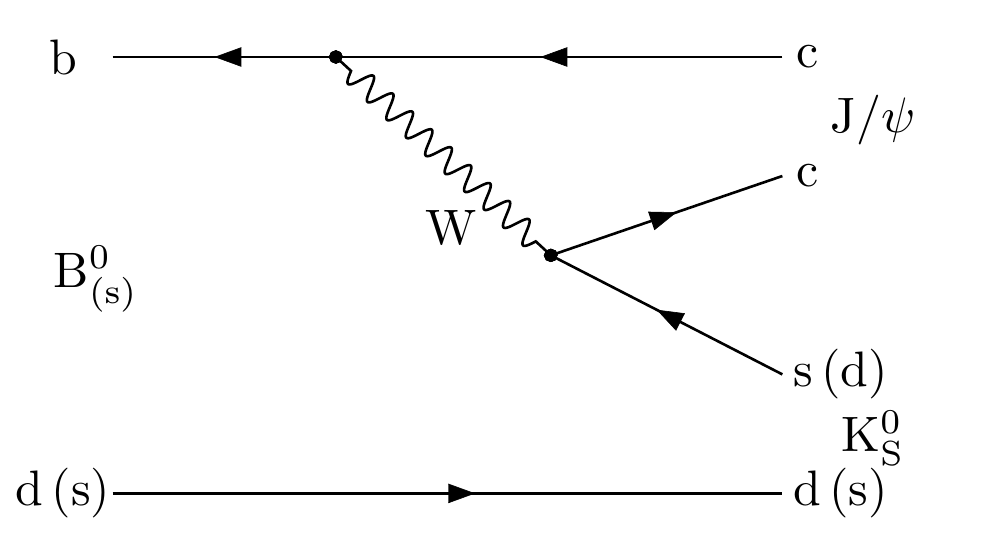}
  \includegraphics[width=0.52\textwidth]{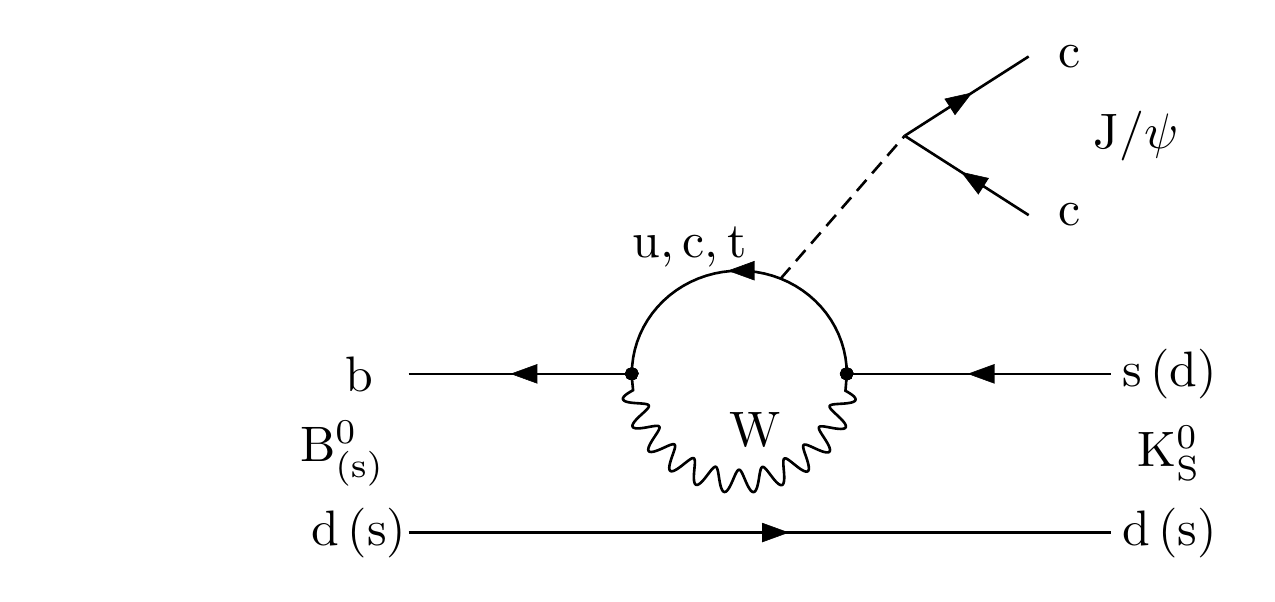}
    \caption{The tree-level (left) and penguin (right) Feynman diagrams for the decays \bdjpsiks and \bsjpsiks.}
  \label{fig:feynman}
\end{figure}

In this paper, we measure the effective lifetime of the \bs meson in its decay to \jpsiks.
Charge conjugate states and decays are implied unless otherwise indicated.
Figure~\ref{fig:feynman} shows the tree-level and penguin Feynman diagrams for the decay modes \bdjpsiks and \bsjpsiks.
The effective lifetime is theoretically defined as the time expectation value of the untagged decay rate~\cite{De_Bruyn_2012}, where 
untagged means that no requirements are made that would bias one \bs mass 
eigenstate over the other. Typically, an untagged rate is characterized by two exponentials, corresponding to the two mass eigenstates with distinct lifetimes,
while a single exponential fit serves as an approximation. The CMS Collaboration 
has measured the effective lifetime in the decay \bmm~\cite{2023137955, CMS:2019bbr}, 
a \cp-odd final state, as well as other \bs meson decay channels~\cite{CMS:2017ygm}. 
The ATLAS and LHCb experiments have also measured the effective lifetimes for a variety of decay modes~\cite{ATLAS:2023trk, LHCb:2022aql, LHCb:2021vsc, LHCb:2021awg, LHCb:2017rmj, 
LHCb:2016crj, LHCb:2014bqh, LHCb:2014qsd, LHCb:2013cca, LHCb:2013dzm, LHCb:2012zwr, PhysRevLett.109.152002, 2012393}.
The effective lifetime of a particular decay channel can be experimentally determined by fitting a single exponential function to its untagged rate.
The \jpsiks state can be considered a pure \cp-odd state~\cite{PhysRevD_hflav}, with \cp violation in the kaon system being neglected~\cite{CDF:2011fhd}. This allows us to accurately measure the \bs heavy-state lifetime.

The \bsjpsiks effective lifetime is defined as
\begin{equation}
  \tau(\bsjpsiks) \equiv \frac{\int_{0}^{\infty} t \, \{\Gamma[\bs(t) \to \jpsiks ] + \Gamma[\bsb (t) \to \jpsiks]\} \,\rd t}
                              {\int_{0}^{\infty} \, \{\Gamma[\bs(t) \to \jpsiks ] + \Gamma[\bsb (t) \to \jpsiks]\} \,\rd t},
\end{equation}
where $t$ is the proper decay time of the \bs meson.
The effective lifetime can also be expressed as~\cite{Fleischer:2011cw}
\begin{equation}
\tau(\bsjpsiks) =\frac{\tau_{\bs}}{1-y_\PQs^2} \left(\frac{1+2\,{y_{\PQs}} {A}_{\Delta\Gamma} + y_\PQs^2}{1+{y_{\PQs}}{A}_{\Delta\Gamma}}\right),
\label{eq:3}
\end{equation}  
where $A_{\Delta\Gamma}$ is the mass eigenstate rate asymmetry,
$y_\PQs$ is the normalized decay width difference, defined as $y_\PQs = \tau_{\bs} \Delta\Gamma / 2$, $\Delta\Gamma$ is the decay width difference 
between the heavy and light \bs mass eigenstates, and $\tau_{\bs}$ denotes the mean \bs lifetime. 
Using the most recent measurements~\cite{PDG24, PhysRevD_hflav} of these observables, the SM prediction for the \bsjpsiks effective lifetime is
$1.62 \pm 0.02\ps$, where the uncertainty is derived from the uncertainties in the observables in Eq.~\eqref{eq:3}.
The only previous measurement of this effective lifetime is from the LHCb Collaboration, which reported
$\tau(\bsjpsiks) = 1.75 \pm 0.12\stat\pm 0.07\syst\ps$ from data collected in 2011~\cite{LHCb:2013dzm}.
The analysis described in this paper uses two variables, the \jpsiks invariant mass and proper 
decay time, to extract the effective lifetime. The decay \bdjpsiks is used as a control channel for various bias tests and validations.

\section{The CMS detector}

The central feature of the CMS detector is a 6\unit{m} internal diameter superconducting solenoid that provides a
magnetic field of 3.8\unit{T}. Within this solenoid volume are a silicon pixel and strip tracker, a lead tungstate
crystal electromagnetic calorimeter, and a brass and scintillator hadron calorimeter. Each of them is composed of a 
barrel and two endcap sections. 
The silicon tracker used in 2016 measured charged particles within the pseudorapidity
range $\abs{\eta} < 2.5$. For nonisolated particles of transverse momentum $1 < \pt < 10\GeV$ 
and $\abs{\eta} < 1.4$, the track resolutions were typically 1.5\% in \pt 
and 25--90 (45--150)\mum in the transverse (longitudinal) impact 
parameter~\cite{CMS:2014pgm}. At the start of 2017, a new pixel detector 
was installed~\cite{Phase1Pixel}; the upgraded tracker measured particles 
up to $\abs{\eta} < 3.0$ with typical resolutions of 1.5\% in \pt and 
20--75\mum in the transverse impact parameter~\cite{DP-2020-049} for 
nonisolated particles of $1 < \pt < 10\GeV$.
Muons with $\abs{\eta} < 2.4$ are measured with gas-ionization chambers embedded in the steel flux-return yoke that 
resides outside the solenoid. A more detailed description of the CMS detector, definition of the coordinate system used, 
and the other relevant kinematic variables, can be found in Ref.~\cite{CMS}. 

Events of interest are selected using a two-tiered trigger system. The 
first level, composed of custom hardware processors, uses information from 
the calorimeters and muon detectors to select events at a rate of around 
100\unit{kHz} within a fixed latency of 4\mus~\cite{CMS:2020cmk}. The 
second level, known as the high-level trigger (HLT), consists of a farm of 
processors running a version of the full event reconstruction software 
optimized for fast processing, and reduces the event rate to around 
1\unit{kHz} before data storage~\cite{CMS:2016ngn}. 
\section{Data and simulation samples}
Data used in the analysis correspond to an integrated luminosity of 140\fbinv~\cite{CMS-LUM-17-003, 
CMS-PAS-LUM-17-004, CMS-PAS-LUM-18-002}, collected by the CMS experiment during 2016--2018 in proton-proton ($\Pp\Pp$) collisions at a 
center-of-mass energy of 13\TeV.

The analysis uses Monte Carlo (MC) samples to estimate the detector acceptance and the trigger and 
reconstruction efficiencies. The MC event samples are simulated using \PYTHIA v8.240~\cite{Pythia8}
and the unstable particle decays are modeled by \EVTGEN~v1.6.0~\cite{EvtGen}. Final-state radiation is included in \EVTGEN using
\PHOTOS~v3.61~\cite{photos}. The particles are propagated through a detailed model of the CMS detector using \GEANTfour~\cite{Geant4}. 
The simulated events follow the same reconstruction algorithm and selection criteria as the collision 
data. To take into account the effect of additional $\Pp\Pp$ collision vertices in the same 
or nearby bunch crossings, the number and spatial distribution of collision vertices in simulated events 
are matched to those from data.

\section{Event selection}

The events used in this analysis were collected by an HLT path that required two oppositely charged muons with a distance
of closest approach less than 0.5\cm, a vertex fit \chisq probability greater than 0.5\%, and an invariant mass in the range 
2.9--3.3\GeV. The muons were required to pass within 2\mm of the beamline and the dimuon system was required to have 
$\abs{\eta} < 2.5$ and $\pt > 19.9\GeV$ in 2016 and $\pt > 24.9\GeV$ in 2017--2018. 
These requirements are also applied on the corresponding offline quantities, with an additional requirement of the muon 
candidates satisfying the soft muon identification criteria~\cite{CMS:2018rym}.

The \jpsiks candidates are reconstructed using the \jpsi meson decay to \mumu 
and \ks meson decay to \pipi. 
After the standard track selection criteria~\cite{Khachatryan:2010pw}, the two daughter tracks assumed to be from the \ks 
decay are given charged pion mass assignments and fit to a common vertex.
The fitted two-track vertex must have a $\chisq<7$~\cite{Kalman} and be located at a 
distance from the beamline that is at least 15 times larger than its uncertainty. The \pipi and \mumu invariant mass
distributions are each fit with a sum of two Gaussian functions with a common mean for the signal component and an exponential function for the background.
The effective invariant mass resolutions are evaluated as $\sigma_{\text{eff}} = \sqrt{\smash[b]{f \sigma_{1}^2 + (1-f) \sigma_{2}^2}}$, where 
$\sigma_1$ and $\sigma_2$ are the individual widths of the two Gaussian functions and $f$ is the 
fraction of events associated with the first Gaussian function.
The effective mass resolutions are approximately 5.5 and 30\MeV for the \ks and \jpsi, respectively.
The \pipi and \mumu candidates are then required to have invariant masses within $2.5\,\sigma_{\text{eff}}$ from 
the corresponding common mean value found in the fit.
The contamination in the \ks candidates from \ltoprpi 
decay, where the proton track is reconstructed as 
a pion, is removed using a requirement on the momentum asymmetry between the two daughters with the Armenteros--Podolanski method~\cite{podolanski1954}.

The two tracks from the $\ks \to \pipi$ candidate decay are fit to a common
vertex with the invariant mass constrained to the \ks world-average mass value~\cite{PDG24}.
The \PB meson candidates are obtained from a vertex fit of \ks\mumu, with the dimuon mass constrained to the world-average \jpsi meson mass.
The refitted momenta are then used to determine the reconstructed invariant mass $m$
of the \PB candidate, which is required to satisfy $4.9<m<6.0\GeV$.

The distance between the \PB and \ks decay vertices in the transverse plane, referred to as the \ks decay length, is required to be 
greater than 5 times its uncertainty. This is done to further suppress background events that do not contain genuine \ks decays, such 
as $\PBz \to \jpsi \Kstar$ where the \Kstar decays to $\PKp\PGpm$ and the \PKp is reconstructed as a \PGpp.

The measurement of the \PB meson decay length requires a precise reconstruction of both 
the production and decay vertices of the \PB meson. 
The proper decay time $t$ of the \PB meson candidate is determined using
 \begin{equation}
      t = m \frac{\vec{L}_{xy}\cdot\vec{p}_\mathrm{T}^{\,\PB}}{\left(p_\text{T}^\text{B}\right)^2},
      \label{dt}
\end{equation}
where $m$ and $\vec{p}_\mathrm{T}^{\,\PB}$ are the invariant mass and transverse momentum of the \PB candidate, respectively, and $\vec{L}_{xy}$ 
is the vector in the transverse plane from the beamline to the \PB meson candidate fitted decay vertex.
The uncertainty in the proper decay time is determined by propagating the
uncertainties in the decay length, invariant mass, and momentum of the
\PB candidate.

To reduce contributions from non-signal processes, a multivariate selection is employed, based
on a boosted decision tree (BDT). 
This BDT relies on 8 discriminating variables, which are associated with the 
properties of the \PB meson candidate and the final-state particles. 
Variables with minimal correlation to the proper decay time are used. 
These variables are: 
the angle between $\vec{L}_{xy}$ and the $\vec{p}_\mathrm{T}^{\,\PB}$;
the \PB and \ks vertex \chisq fit probabilities; the \pt of the \jpsi and \ks candidates;
the $\eta$ of the \PB candidate; the angle in the transverse plane between the \ks momentum vector and 
the line joining the \PB and \ks candidate decay vertices; 
and the impact parameter of the \PB candidate with respect to the beamline divided by its uncertainty.

The BDT classifier uses two input samples containing signal and background decays, respectively.
The signal sample is selected from MC events, and the background sample is made up
of data events with the \PB meson candidate invariant mass in the high sideband region of
5.6--6.0\GeV. A distinct BDT is used for each of the three data-taking years to account for
variations in the data-simulation agreement and different detector
conditions.
The optimization of the BDT threshold is done by minimizing the statistical uncertainty in the effective lifetime measurement based on studies using pseudo-experiments.
In events with more than one \PB meson candidate satisfying the signal requirements, the one with the
highest BDT discriminant value is chosen. To validate the reliability of the BDT approach, different kinematic 
distributions are compared and found to be consistent between the MC samples and the background-subtracted data for the control channel decay 
\bdjpsiks.

\section{Effective lifetime measurement}

The effective lifetime for the signal \bs meson is extracted using a single two-dimensional unbinned extended maximum 
likelihood (2D UML) fit to the \jpsiks invariant mass ($m$) and the proper decay time ($t$) distributions, with the calculated uncertainty 
in the proper decay time ($\st$) as a conditional parameter in the fit. This fit is performed simultaneously 
to each of the three data-taking years, with independent parameters for each year except for the effective lifetimes of the signal and 
control channels, which are each represented by a single parameter. The decay time and invariant mass range used in the fit is 
$0.2 < t < 10\ps$ and $5.17 < m < 5.57\GeV$, respectively, with the mass range not extended below 5.17\GeV to avoid contributions 
from partially reconstructed decays of \Pb hadrons. The signal efficiency as a function of decay time is determined from simulated samples and 
is used to account for the distortion of the measured decay time distribution caused by the selection criteria.

The unnormalized probability density function (pdf) $\mathcal{L}$ is expressed as
  \begin{equation}
  \begin{split}
  \mathcal{L}(m, t;\, \st) = N_{\bs}\, M_{\bs}(m)\, T_{\bs}(t; \,\st)\, \varepsilon_{\bs}(t)  
  + N_{\bd}\, M_{\bd}(m) \,  T_{\bd}(t; \,\st) \:\varepsilon_{\bd}(t)\\
  +\, N_{\text{bkg}} \, M_{\text{bkg}}(m) \, T_{\text{bkg}}(t; \,\st). \\
  \end{split}
  \end{equation}
Here, $M(m)$ and $T(t;\,\st)$ are the invariant mass and decay time pdfs, respectively. 
The pdfs for the signal, control channel, and background components are labeled with the subscripts \bs, \bd, and \text{bkg}, respectively. 
The signal and control channel efficiencies are given by $\varepsilon_{\bs}$ and 
$\varepsilon_{\bd}$, respectively.
The parameters $N_{\bs}$, $N_{\bd}$, and $N_{\text{bkg}}$ correspond to the signal, control channel, and background yields, respectively. 

The invariant mass distribution from simulated \bdjpsiks events is fit with a sum of three 
Gaussian functions with a common mean. This fitted function, with the two parameters governing
the relative amplitude of each Gaussian fixed, is used in fitting the data. Before fitting the 
data, the function is fit to simulated \bsjpsiks events with the yield, mean, and a single scale
factor (\bs MC) multiplying each of the widths as the free parameters. In fitting the data, the 
two invariant mass signal functions for \bd and \bs have a total of four free parameters: two yields, one mean, 
and one width scale factor. The world-average value for the difference in masses of the \bs and \bd mesons of 
$87.26 \pm 0.24\MeV$~\cite{PDG24} is applied as a Gaussian constraint to the difference of the means of the two 
corresponding invariant mass functions. The width scale factor multiplies
all of the width parameters of each function, with the \bs signal function widths also multiplied by the \bs MC scale factor.
The shape of the invariant mass distribution for the background is parametrized
by an exponential function with a free slope parameter. 

The pdfs $T_{\bs}$ and $T_{\bd}$ are exponential functions 
convolved with a Gaussian function to account for the detector 
resolution. These are multiplied by their respective efficiency functions $\varepsilon_{\bs}(t)$
and $\varepsilon_{\bd}(t)$. A sum of two exponential functions is used to model $T_{\text{bkg}}$, with all parameters left free in the fit.

\begin{figure}[!ht]
  \centering
    \includegraphics[width=0.32\textwidth]{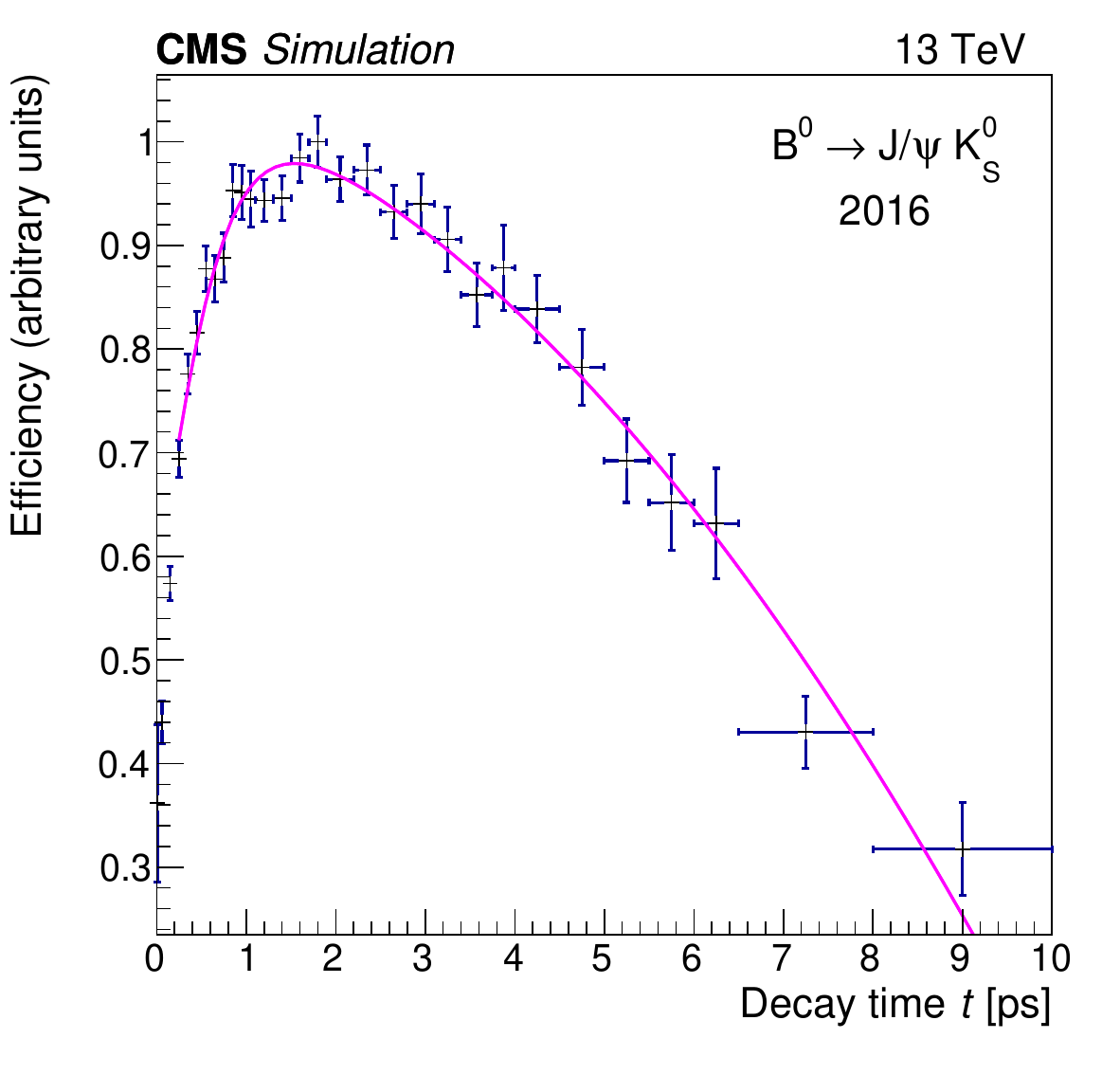}
    \includegraphics[width=0.32\textwidth]{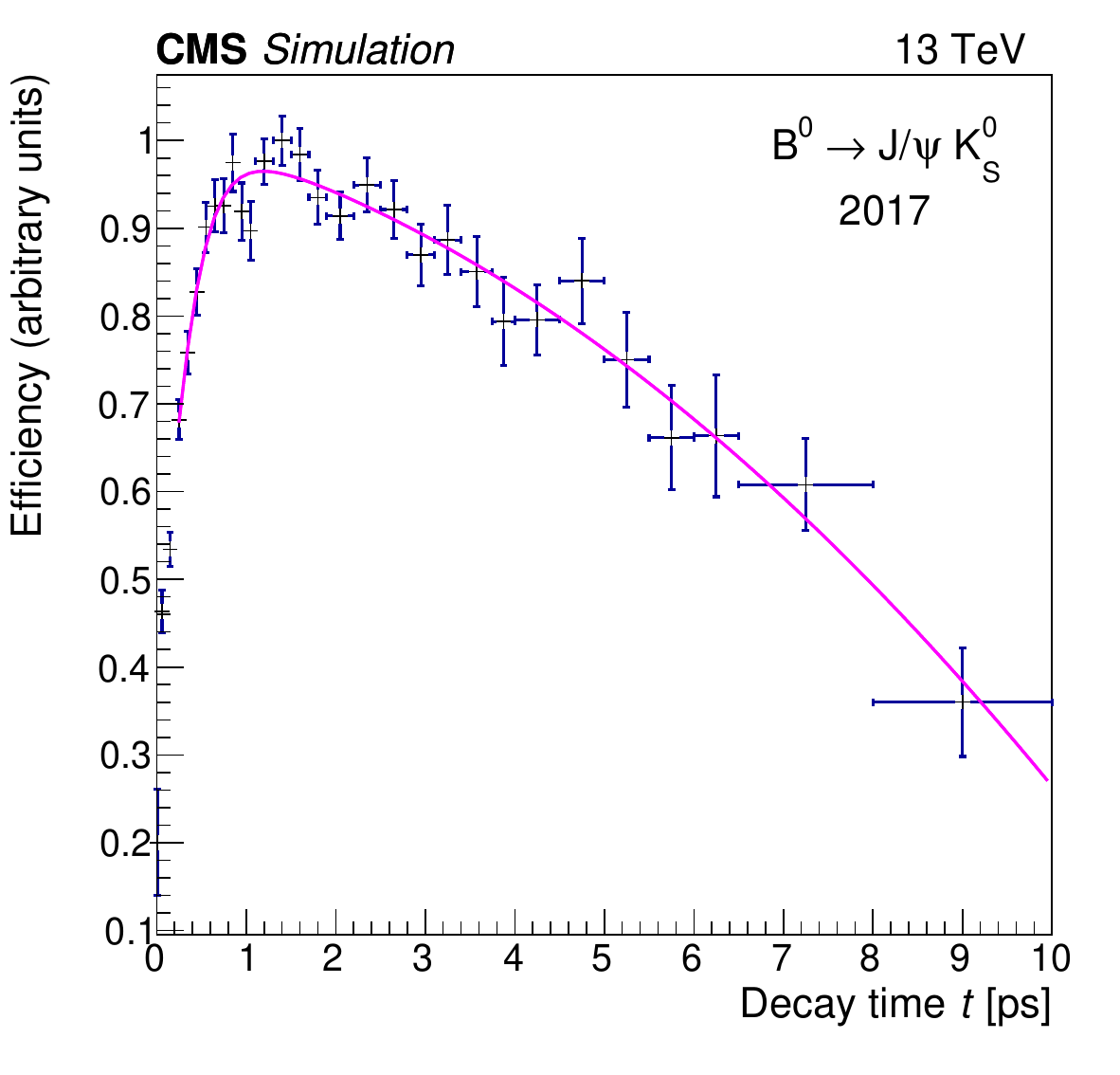}
    \includegraphics[width=0.32\textwidth]{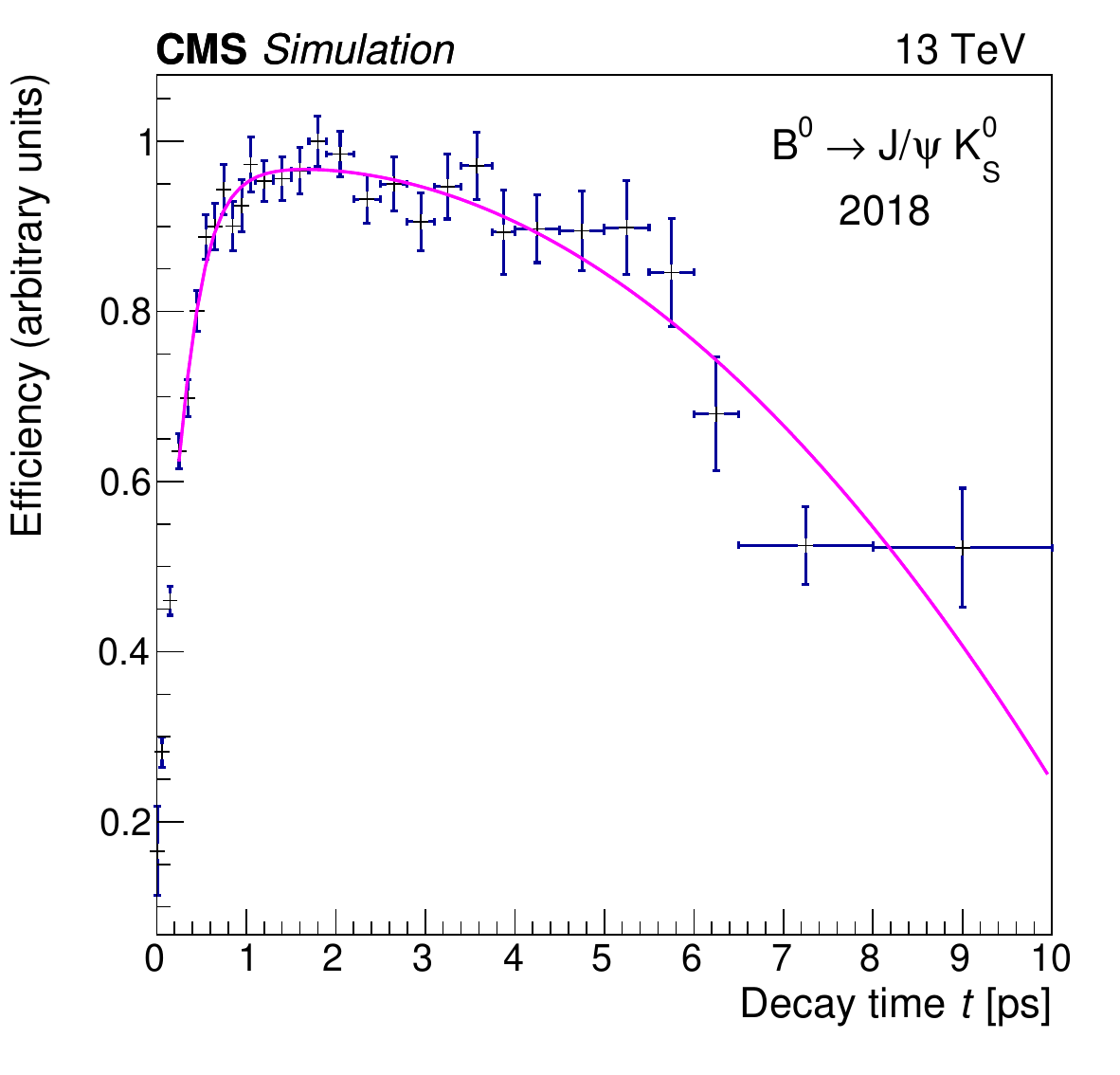}
    \includegraphics[width=0.32\textwidth]{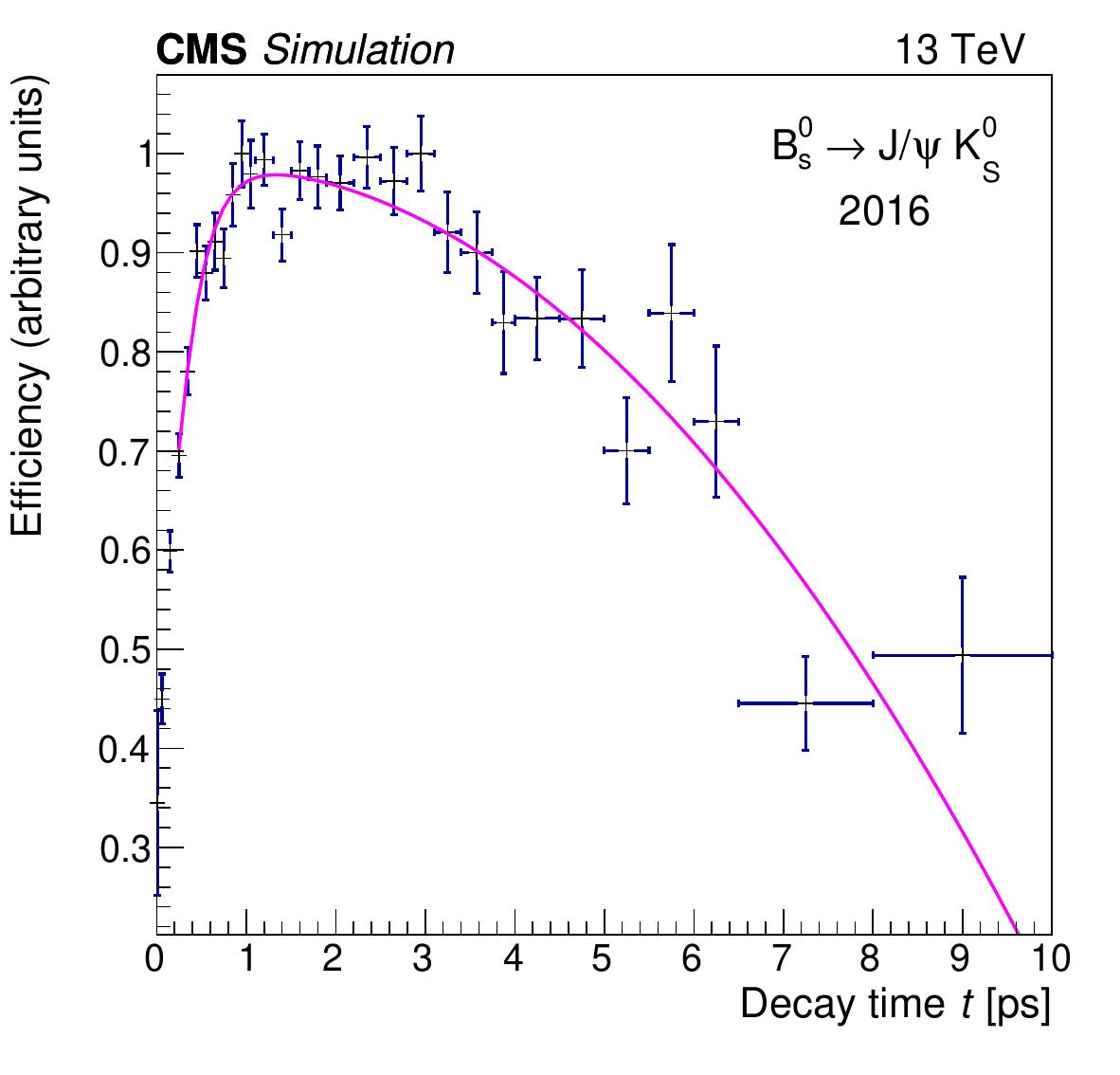}
    \includegraphics[width=0.32\textwidth]{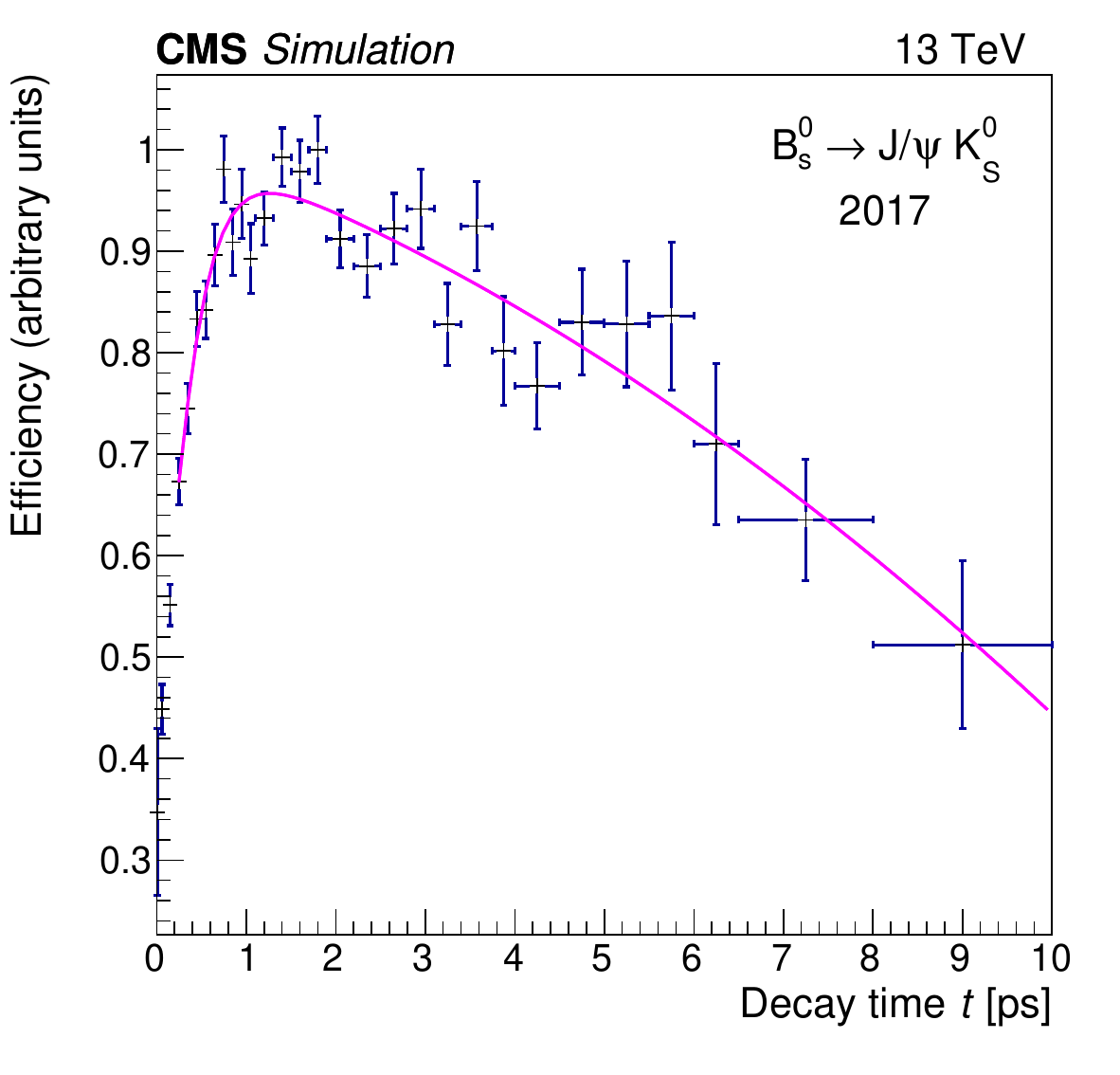}
    \includegraphics[width=0.32\textwidth]{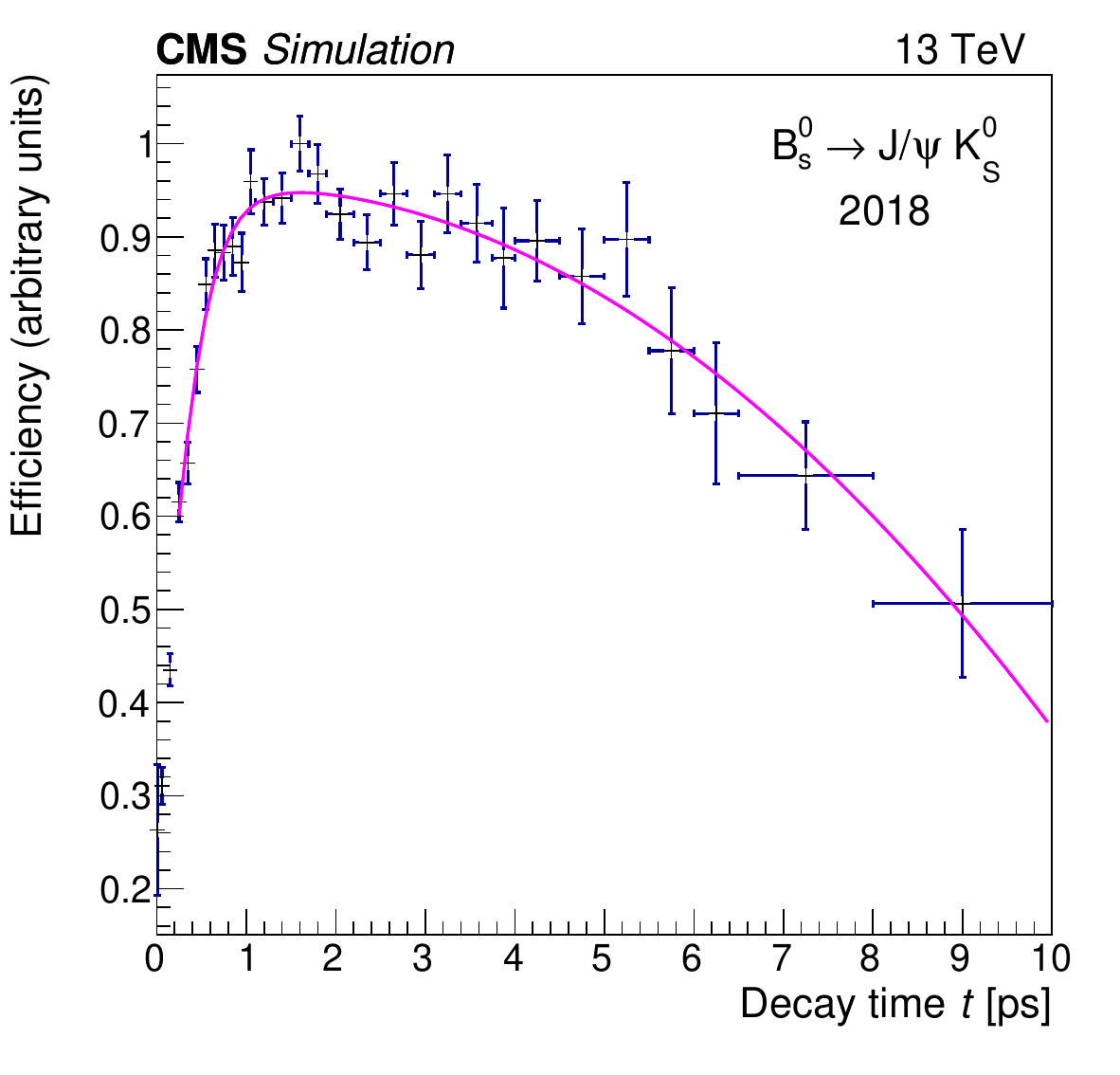}
\caption{The signal efficiency as a function of the decay time for the \bdjpsiks (upper) and \bsjpsiks (lower) decays
    from simulation for each of the three data-taking years.
    The vertical bars indicate the statistical uncertainty, and the horizontal bars give 
    bin widths. The curves show the fit results.}
\label{eff}
\end{figure} 

The efficiencies are obtained by dividing the distribution of the reconstructed 
decay time in the MC by the distribution of the generated decay time convolved
with the time resolution function and are fitted with the function:
\begin{equation}\label{eq:efffunc}
  \varepsilon(t;\,p_{0},p_{1},p_{2},p_{3},p_{4})= p_{0}+p_{1}t + p_{2}t^{2} + \frac{p_{3}}{1+\exp(-p_{4}t)},
\end{equation}
where $p_i$ ($i$ = 0--4) are free parameters in the
fit. The efficiencies, along with the fitted $\varepsilon$ functions, are shown
in Fig.~\ref{eff} for each of the three data-taking years. 
The ratio of the two efficiencies for each year is consistent with unity and shows no dependence on the decay time.

\begin{figure}[!ht]
\centering
    \includegraphics[width=0.48\textwidth]{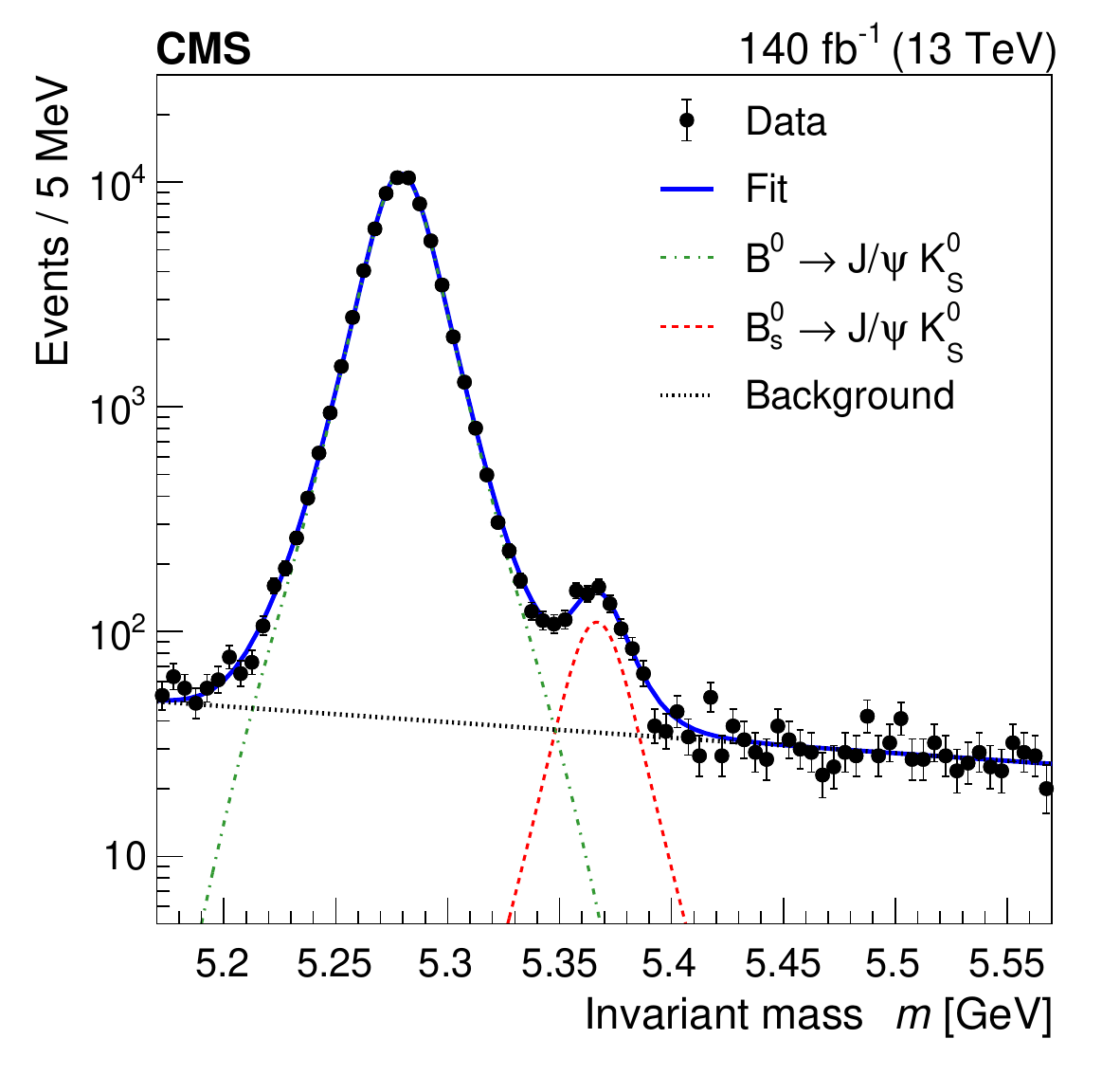}
    \includegraphics[width=0.48\textwidth]{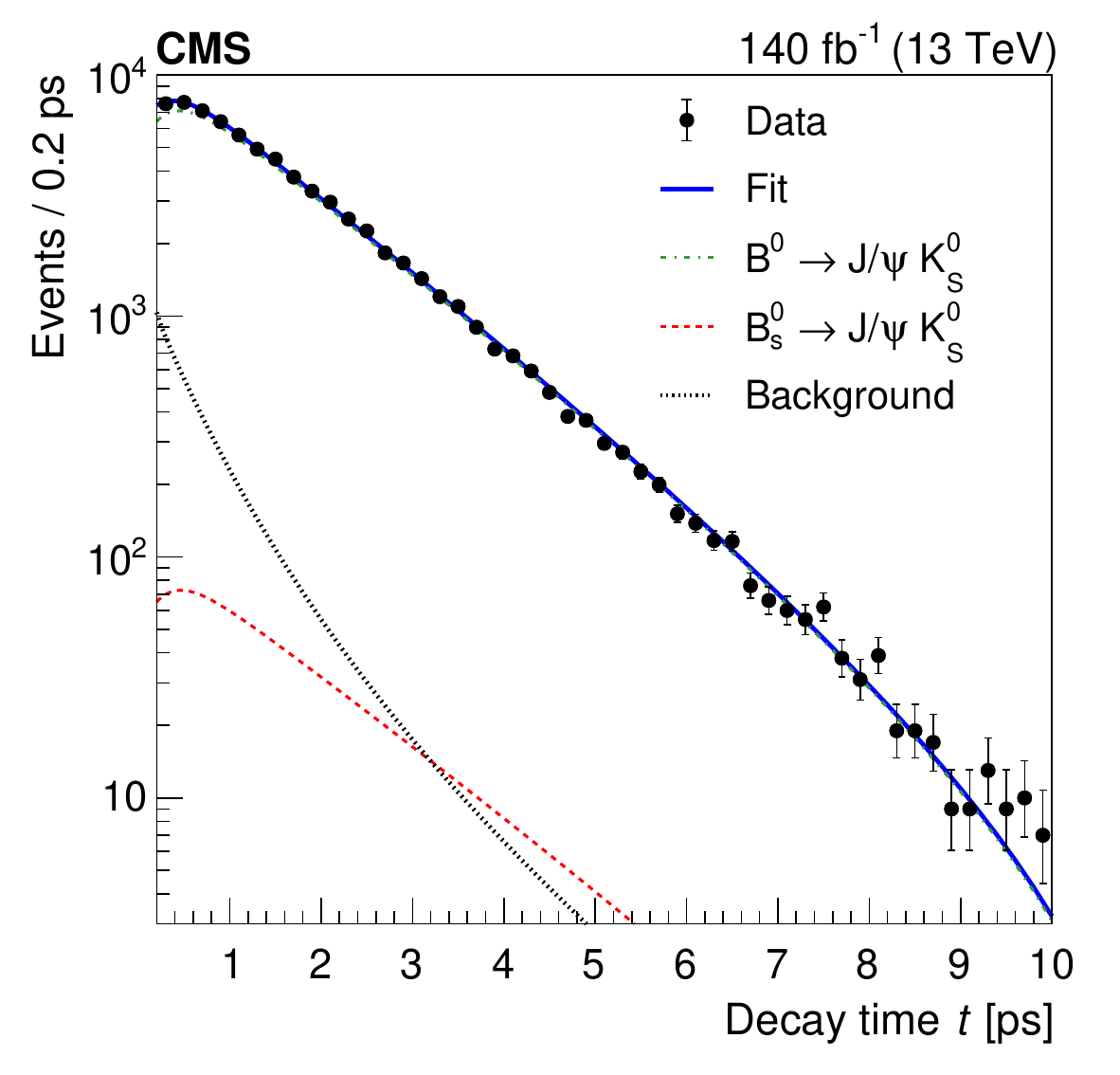}
\caption{Distributions of the \jpsiks invariant mass (left) and
    proper decay time (right) from data (points) and the results from the
    2D UML fit projections (lines) for the 2016--2018 data set.
    The vertical bars on the data points indicate the statistical uncertainty. The solid, dotted-dashed, dashed, and dotted lines
    show the total fit, \bd control channel, \bs signal, and background
    contributions, respectively.}
\label{fig:001}
\end{figure}

\begin{figure}[!ht]
\centering
\includegraphics[width=0.48\textwidth]{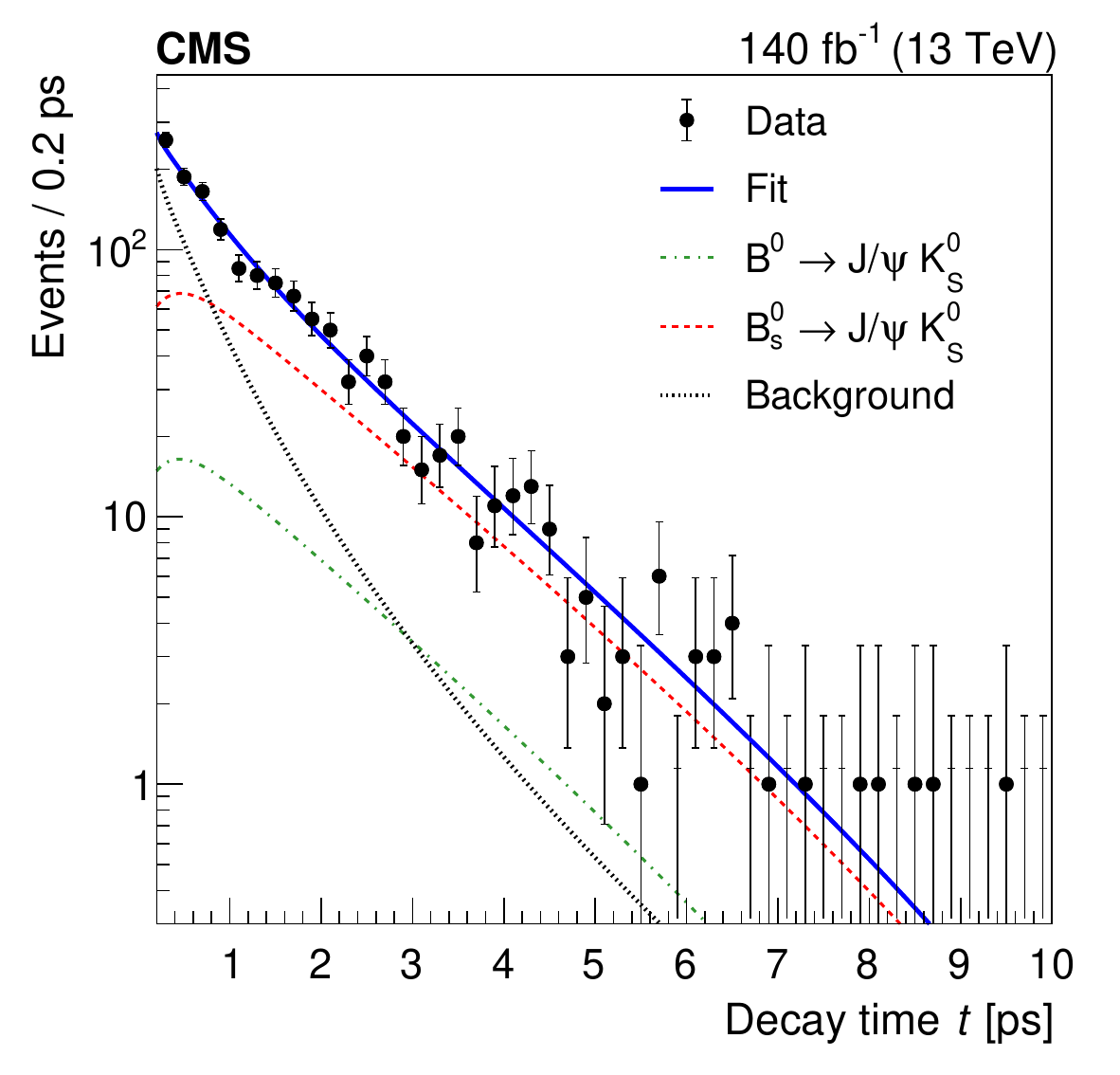}
\caption{The proper decay time distribution from data (points) for
    events in the \bs signal region with \jpsiks invariant mass in the range
    5.34--5.42\GeV and the results from the
    2D UML fit projections (lines) for the 2016--2018 data set. The vertical bars on the data points indicate the
    statistical uncertainty. The dashed, dotted-dashed, dotted, and solid lines
    show the \bs signal, \bd control channel, background, and total fit
    contributions, respectively.}
\label{fig:002}
\end{figure}

The measured $m$ and $t$ distributions are 
shown in Fig.~\ref{fig:001}, along with the projections of the 2D UML fit, in the full fit range and summed over all three data-taking years. 
The invariant mass projection from the 2D UML fit yields a \chisq value of 82 with 80 bins,
while the decay time projection results in a \chisq value of 62 with 49 bins.
For a better visualization of the signal component, the proper decay time projection is shown in Fig.~\ref{fig:002} for events in the signal
invariant mass range $5.34 < m < 5.42\GeV$, along with individual and total pdf
components. The measured effective lifetime for the \bs from the 
simultaneous fit is $\tau(\bsjpsiks) = 1.59 \pm 0.07\stat \ps$.
Projections in $m$ and $t$ of this fit are given for each of the three data-taking years in Appendix~\ref{final_proj}, along with the corresponding data results.

Consistency checks are conducted, such as examining the decay time projections in different 
\bs invariant mass ranges and the invariant mass projections for different decay time ranges. 
There is good agreement between the data and results from the fit in all cases.
These projections are shown in Appendix~\ref{subrange_proj}. 

\section{Systematic uncertainties}
Several sources of systematic uncertainty are investigated and their effects on the \bs
effective lifetime measurement are estimated.

The finite number of simulated events introduces an uncertainty due to the statistical
variation in the MC samples and its effect on the determination of the efficiency functions. 
This is studied by generating 400 alternative efficiency functions, produced by varying the
fitted parameters of the default efficiency function by their uncertainties. Each of these
functions is then used in the fit to data. The standard deviation of the resulting 
effective lifetime measurements from these fits is 0.006\ps, which is used as an estimate of the
systematic uncertainty from this source.

Another source of uncertainty comes from the choice of the function used to parametrize the efficiency as a function of the decay time. 
To estimate this effect, two other analytical functions (variations of Eq.~\eqref{eq:efffunc}) are used for efficiency modeling and each is used in the data fit. 
The largest deviation in the measured value of the effective lifetime from the fit to data using these two functions with respect to the value 
obtained using the default efficiency function is 0.002\ps, which is taken as the uncertainty from this source. 

The choice of pdfs for the modeling of the signal and background invariant mass distributions is another source
of systematic uncertainty. To estimate the effect, alternative functions are used to fit the data. Three alternative signal functions for the invariant mass,
the double-sided Crystal Ball function~\cite{CB}, the Johnson $S_U$ function~\cite{Johnson}, and the Student's \textit{t}-distribution~\cite{Studentt}, are chosen since they provide a good fit to the simulated signal events. 
The alternative background function for the invariant mass is a Bernstein polynomial. 
Additionally, the effect of the uncertainty in the \bs-\bd mass difference evaluated by varying the mass difference in the fits.
The largest variation from the nominal fit, considering all combinations of nominal and alternative functions as well as the different 
mass difference values, is 0.022\ps, which is taken as the systematic uncertainty.

Similarly, the pdf for the decay time distribution of the background is changed to the sum of three exponential functions
and the resulting variation in the result is 0.014\ps.

The \bs MC scale factor that allows the widths of the three Gaussian functions to vary with equal
proportion in the fit is considered as a source of 
uncertainty. The scale factor is changed by its uncertainty (from the fit to the \bsjpsiks MC events) and 
the variation in the \bs effective lifetime is 0.004\ps.

An alternative 2D UML fit is performed in which the ratio of the \bs to \bd signal yields is constrained to 
be the same for all three data-taking years rather than being independent for each year.
The change in the final result of 0.006\ps is taken as the corresponding systematic uncertainty.
It is possible that a small number of reconstructed \bs mesons may come from \bc decays. A fit was performed that added a component with 
the same mass shape as the signal, a lifetime distribution given by an exponential with a time constant equal to the world-average \bc average 
lifetime plus the fitted \bs lifetime, and an unconstrained yield. The difference in \bs lifetime from the nominal fit of 0.002\ps is taken as 
a systematic uncertainty.

The systematic uncertainty from the BDT is negligible; varying the BDT threshold by $\pm5\%$ resulted in consistent lifetime measurements.
The invariant mass range used in the fit is changed from the default of 5.17--5.57 to 5.1--5.6\GeV, with no effect on the result.
To determine the potential uncertainty arising from an inaccurate \st calculation, the values of \st have been scaled by 0.5 and 2 in the fit. 
No change in the lifetime is found and therefore no systematic uncertainty is assessed.

As previously noted, the control channel \bdjpsiks has the same topology and very similar kinematics to those of 
the signal channel, and is therefore used in defining the analysis strategy and to perform various checks. 
As the \bd meson lifetime extracted in this analysis, $1.521 \pm 0.007 \stat \ps$, agrees with the world-average 
value of $1.517 \pm 0.004 \ps$~\cite{PDG24}, we assign an uncertainty of 0.007\ps, based on the precision of this test, 
as an estimate of any unaccounted systematic uncertainties.

All the sources of systematic uncertainty and their estimated values are
listed in Table~\ref{syst}.
The individual values are added in quadrature to obtain the total
systematic uncertainty. 

\begin{table}[hbt]
\centering
  \topcaption{Sources of systematic uncertainties in the \bsjpsiks effective lifetime measurement
and their estimated values, along with the total systematic uncertainty.}
\label{syst}
\begin{tabular}{lc} 
Source                                  & Value (ps)\\ \hline
Limited MC event count                  & 0.006     \\
Efficiency modeling                     & 0.002     \\ 
Signal and background invariant mass model & 0.022  \\
Background decay time model             & 0.014     \\
Invariant mass shape variation          & 0.004     \\
Different fit strategy                  & 0.006     \\ 
Contribution of \bs from \bc decays     & 0.002     \\
Control channel lifetime uncertainty    & 0.007     \\ [\cmsTabSkip]
Total                                   & 0.029 
\end{tabular}
\end{table}

\section{Results}
The 2D UML fit to the complete data set determines the yields of the
\bs signal and \bd control channel to be
$727 \pm 35$ and $68\,460 \pm 270$, respectively, where the
uncertainties are statistical only.
The \bsjpsiks effective lifetime is measured to be:
\begin{equation}
     \tau(\bsjpsiks) = 1.59 \pm 0.07 \stat \pm 0.03 \syst \ps.
\end{equation}

This result is consistent with the expectation from the SM of $1.62 \pm 0.02 \ps$ and is twice as
precise as the only previous measurement of $1.75 \pm 0.12\stat\pm 0.07\syst\ps$,
reported by the LHCb Collaboration from data collected in 2011~\cite{LHCb:2013dzm}. Tabulated results for
this analysis are provided in the HEPData record~\cite{hepdata}.

\section{Summary}
In this paper, a measurement of the effective lifetime of the \bs meson in the \jpsiks decay channel is presented. 
The analysis is performed using data collected by the CMS detector during proton-proton collisions at a 
center-of-mass energy of 13\TeV from 2016 to 2018, 
corresponding to an integrated luminosity of 140\fbinv.
The effective lifetime is extracted using a two-dimensional unbinned maximum likelihood fit to the invariant mass and proper decay 
time distributions of the \bs meson. The decay \bdjpsiks, which has a much larger event yield than the
corresponding \bs decay, is used as a control channel for estimating resolutions
and systematic uncertainties. The measured value of the effective lifetime is 
$\tau(\bsjpsiks) = 1.59 \pm 0.07 \stat \pm 0.03 \syst\ps$, which is the most 
precise result to date.
This measurement can be used to constrain the parameters that govern mixing and \cp violation in the \bs 
system and also to better understand the penguin contributions in measurements of $\sin{(2\beta)}$ from \bdjpsiks decays.

\begin{acknowledgments}
  We congratulate our colleagues in the CERN accelerator departments for the excellent performance of the LHC and thank the technical and administrative staffs at CERN and at other CMS institutes for their contributions to the success of the CMS effort. In addition, we gratefully acknowledge the computing centers and personnel of the Worldwide LHC Computing Grid and other centers for delivering so effectively the computing infrastructure essential to our analyses. Finally, we acknowledge the enduring support for the construction and operation of the LHC, the CMS detector, and the supporting computing infrastructure provided by the following funding agencies: SC (Armenia), BMBWF and FWF (Austria); FNRS and FWO (Belgium); CNPq, CAPES, FAPERJ, FAPERGS, and FAPESP (Brazil); MES and BNSF (Bulgaria); CERN; CAS, MoST, and NSFC (China); MINCIENCIAS (Colombia); MSES and CSF (Croatia); RIF (Cyprus); SENESCYT (Ecuador); ERC PRG, RVTT3 and MoER TK202 (Estonia); Academy of Finland, MEC, and HIP (Finland); CEA and CNRS/IN2P3 (France); SRNSF (Georgia); BMBF, DFG, and HGF (Germany); GSRI (Greece); NKFIH (Hungary); DAE and DST (India); IPM (Iran); SFI (Ireland); INFN (Italy); MSIP and NRF (Republic of Korea); MES (Latvia); LMTLT (Lithuania); MOE and UM (Malaysia); BUAP, CINVESTAV, CONACYT, LNS, SEP, and UASLP-FAI (Mexico); MOS (Montenegro); MBIE (New Zealand); PAEC (Pakistan); MES and NSC (Poland); FCT (Portugal);  MESTD (Serbia); MCIN/AEI and PCTI (Spain); MOSTR (Sri Lanka); Swiss Funding Agencies (Switzerland); MST (Taipei); MHESI and NSTDA (Thailand); TUBITAK and TENMAK (Turkey); NASU (Ukraine); STFC (United Kingdom); DOE and NSF (USA).
   
  \hyphenation{Rachada-pisek} Individuals have received support from the Marie-Curie program and the European Research Council and Horizon 2020 Grant, contract Nos.\ 675440, 724704, 752730, 758316, 765710, 824093, 101115353, 101002207, and COST Action CA16108 (European Union); the Leventis Foundation; the Alfred P.\ Sloan Foundation; the Alexander von Humboldt Foundation; the Science Committee, project no. 22rl-037 (Armenia); the Belgian Federal Science Policy Office; the Fonds pour la Formation \`a la Recherche dans l'Industrie et dans l'Agriculture (FRIA-Belgium); the F.R.S.-FNRS and FWO (Belgium) under the ``Excellence of Science -- EOS" -- be.h project n.\ 30820817; the Beijing Municipal Science \& Technology Commission, No. Z191100007219010 and Fundamental Research Funds for the Central Universities (China); the Ministry of Education, Youth and Sports (MEYS) of the Czech Republic; the Shota Rustaveli National Science Foundation, grant FR-22-985 (Georgia); the Deutsche Forschungsgemeinschaft (DFG), among others, under Germany's Excellence Strategy -- EXC 2121 ``Quantum Universe" -- 390833306, and under project number 400140256 - GRK2497; the Hellenic Foundation for Research and Innovation (HFRI), Project Number 2288 (Greece); the Hungarian Academy of Sciences, the New National Excellence Program - \'UNKP, the NKFIH research grants K 131991, K 133046, K 138136, K 143460, K 143477, K 146913, K 146914, K 147048, 2020-2.2.1-ED-2021-00181, and TKP2021-NKTA-64 (Hungary); the Council of Science and Industrial Research, India; ICSC -- National Research Center for High Performance Computing, Big Data and Quantum Computing and FAIR -- Future Artificial Intelligence Research, funded by the NextGenerationEU program (Italy); the Latvian Council of Science; the Ministry of Education and Science, project no. 2022/WK/14, and the National Science Center, contracts Opus 2021/41/B/ST2/01369 and 2021/43/B/ST2/01552 (Poland); the Funda\c{c}\~ao para a Ci\^encia e a Tecnologia, grant CEECIND/01334/2018 (Portugal); the National Priorities Research Program by Qatar National Research Fund;  MCIN/AEI/10.13039/501100011033, ERDF ``a way of making Europe", and the Programa Estatal de Fomento de la Investigaci{\'o}n Cient{\'i}fica y T{\'e}cnica de Excelencia Mar\'{\i}a de Maeztu, grant MDM-2017-0765 and Programa Severo Ochoa del Principado de Asturias (Spain); the Chulalongkorn Academic into Its 2nd Century Project Advancement Project, and the National Science, Research and Innovation Fund via the Program Management Unit for Human Resources \& Institutional Development, Research and Innovation, grant B39G670016 (Thailand); the Kavli Foundation; the Nvidia Corporation; the SuperMicro Corporation; the Welch Foundation, contract C-1845; and the Weston Havens Foundation (USA).
\end{acknowledgments}

\bibliography{auto_generated}

\appendix
  
\section{Projection plots from the 2D UML fits} \label{final_proj}

\begin{figure}[htb]
  \centering
    \includegraphics[width=0.32\textwidth]{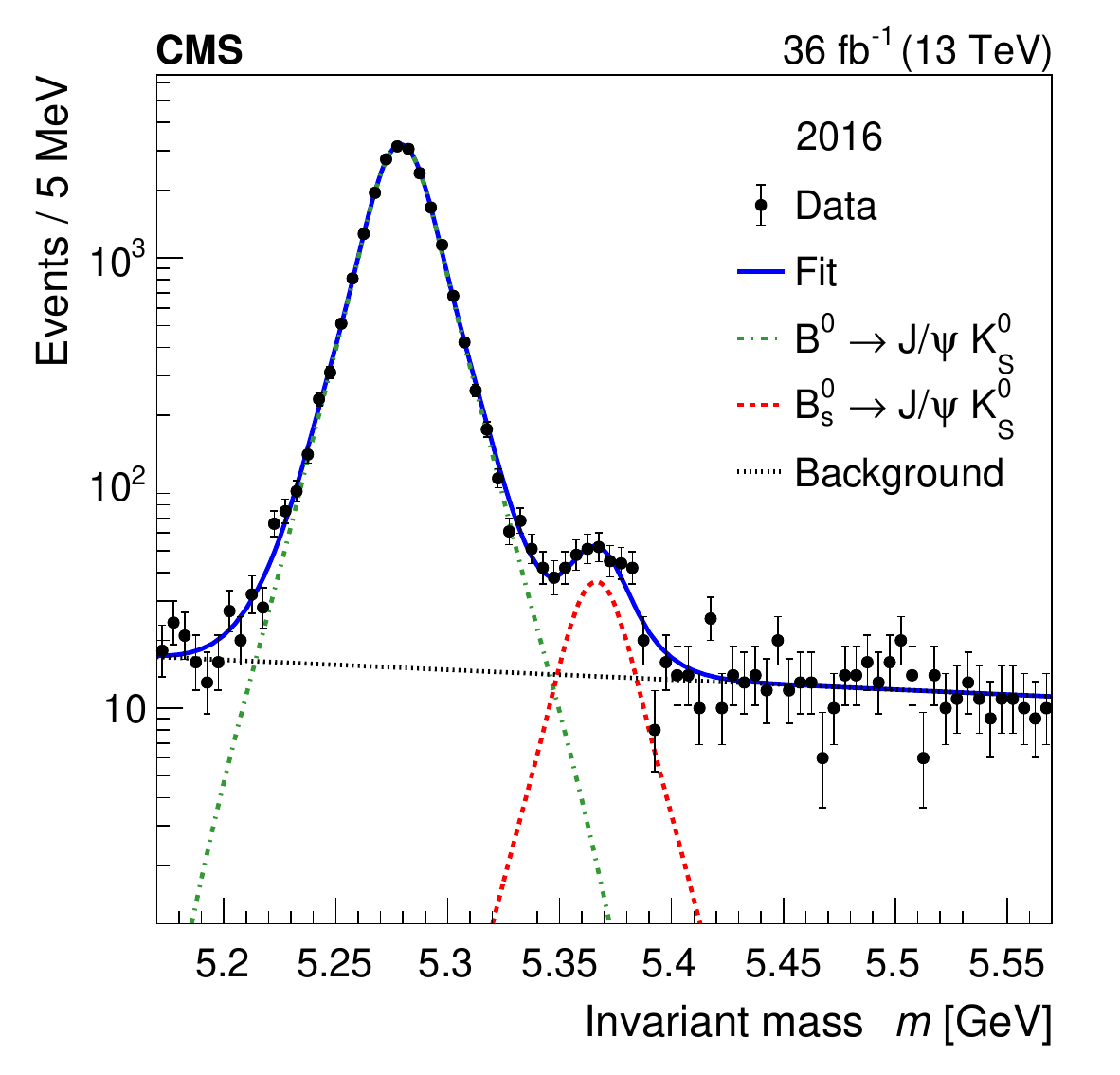}
    \includegraphics[width=0.32\textwidth]{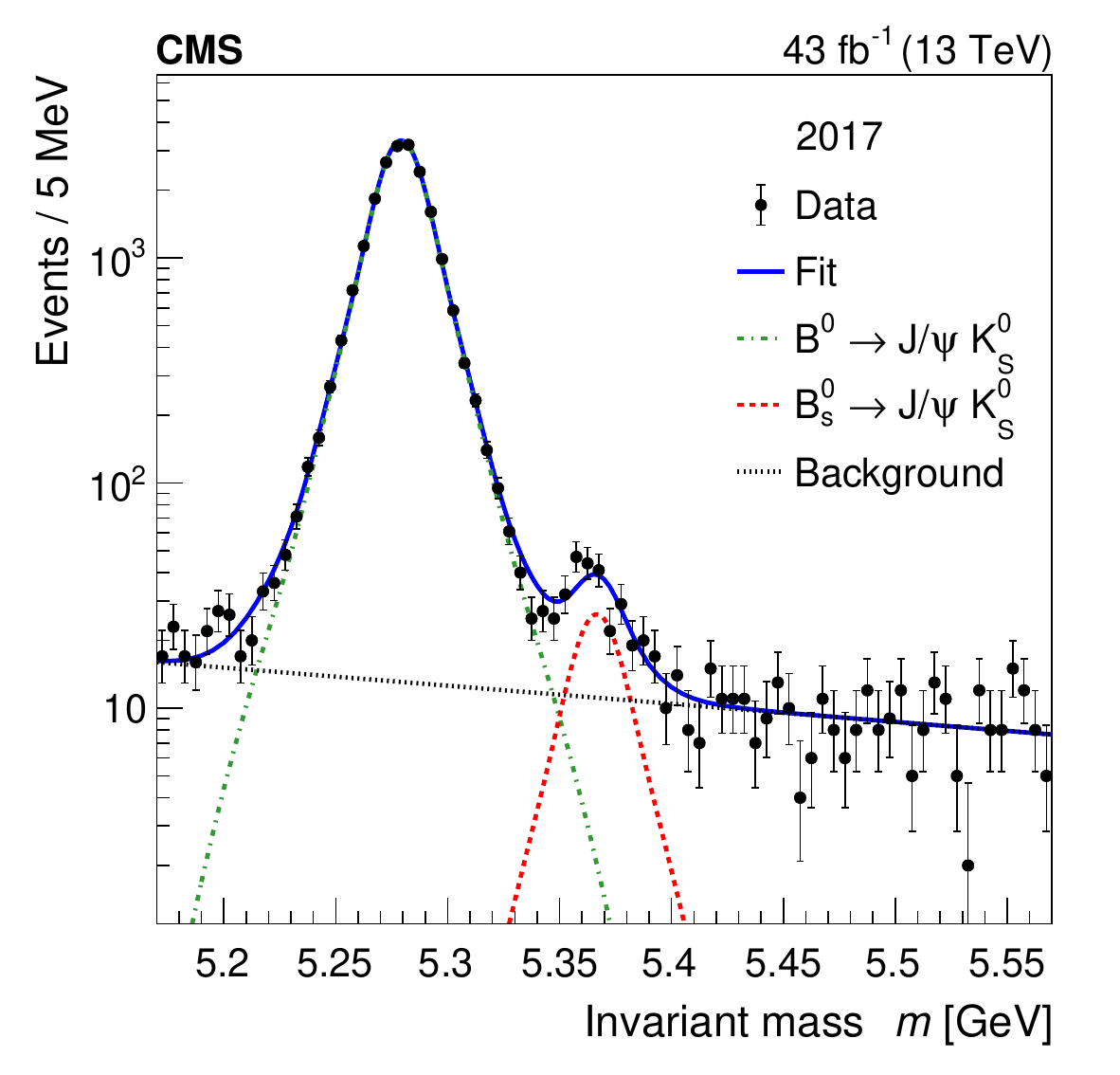}
    \includegraphics[width=0.32\textwidth]{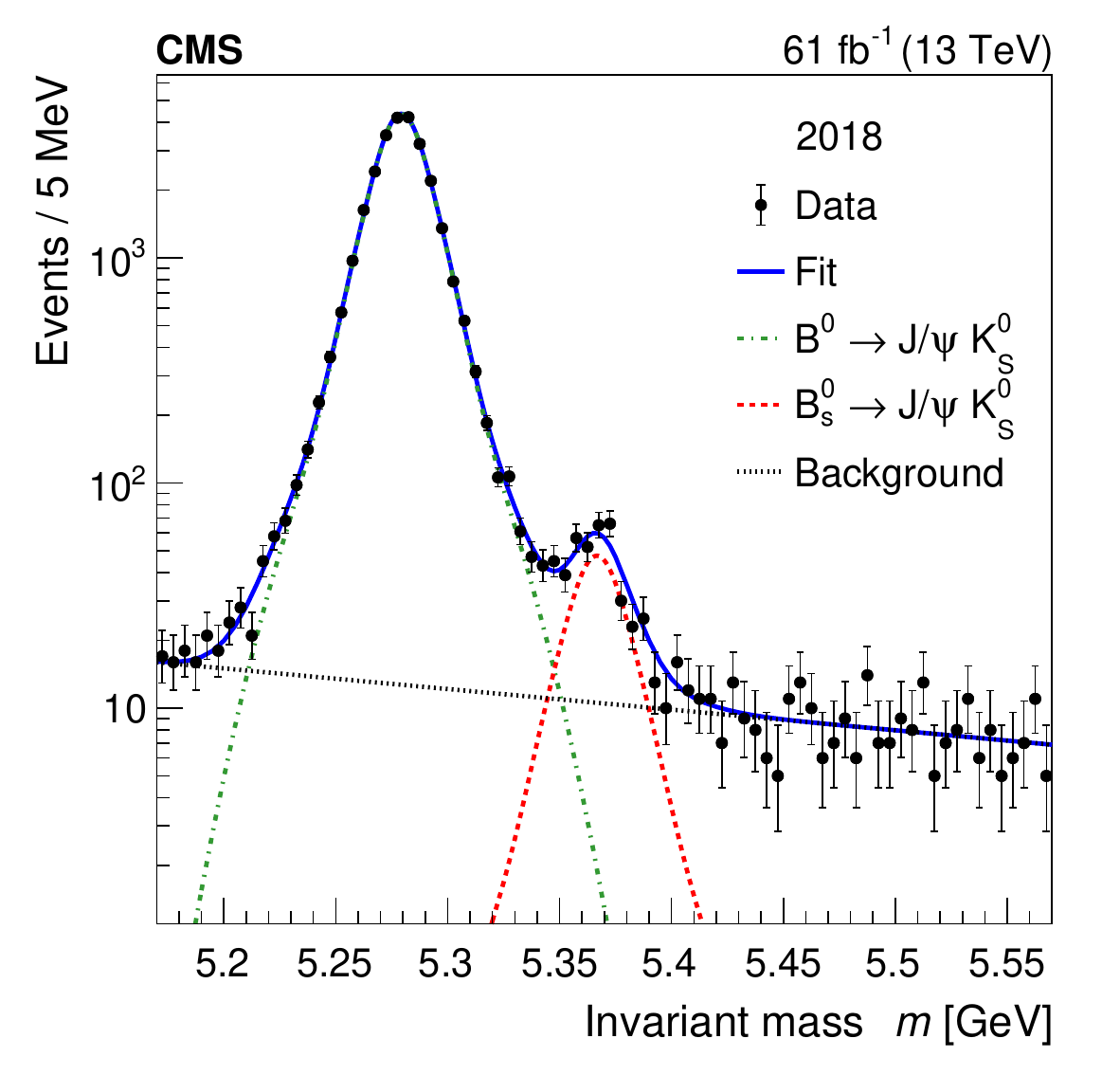}
    \includegraphics[width=0.32\textwidth]{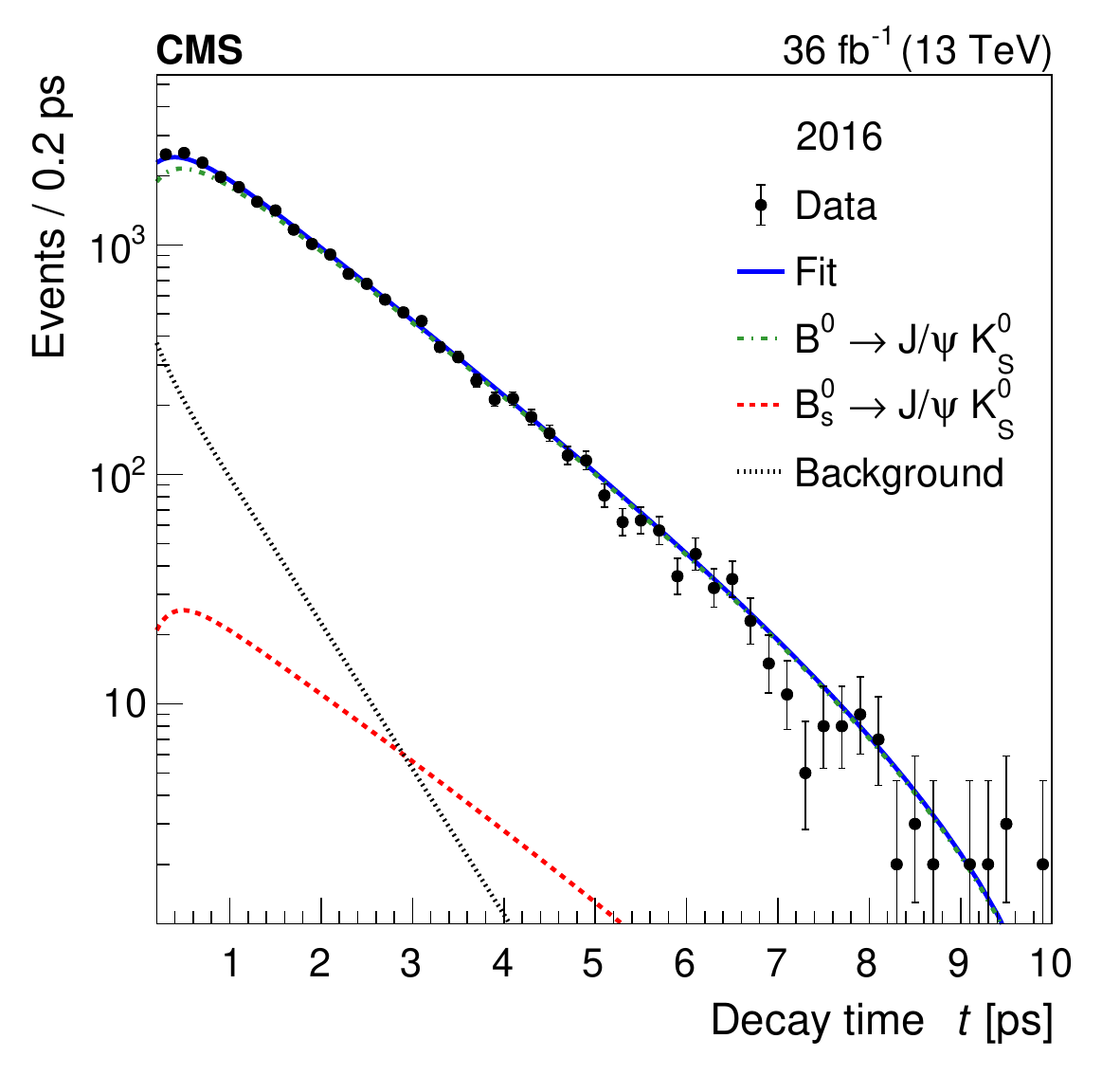}
    \includegraphics[width=0.32\textwidth]{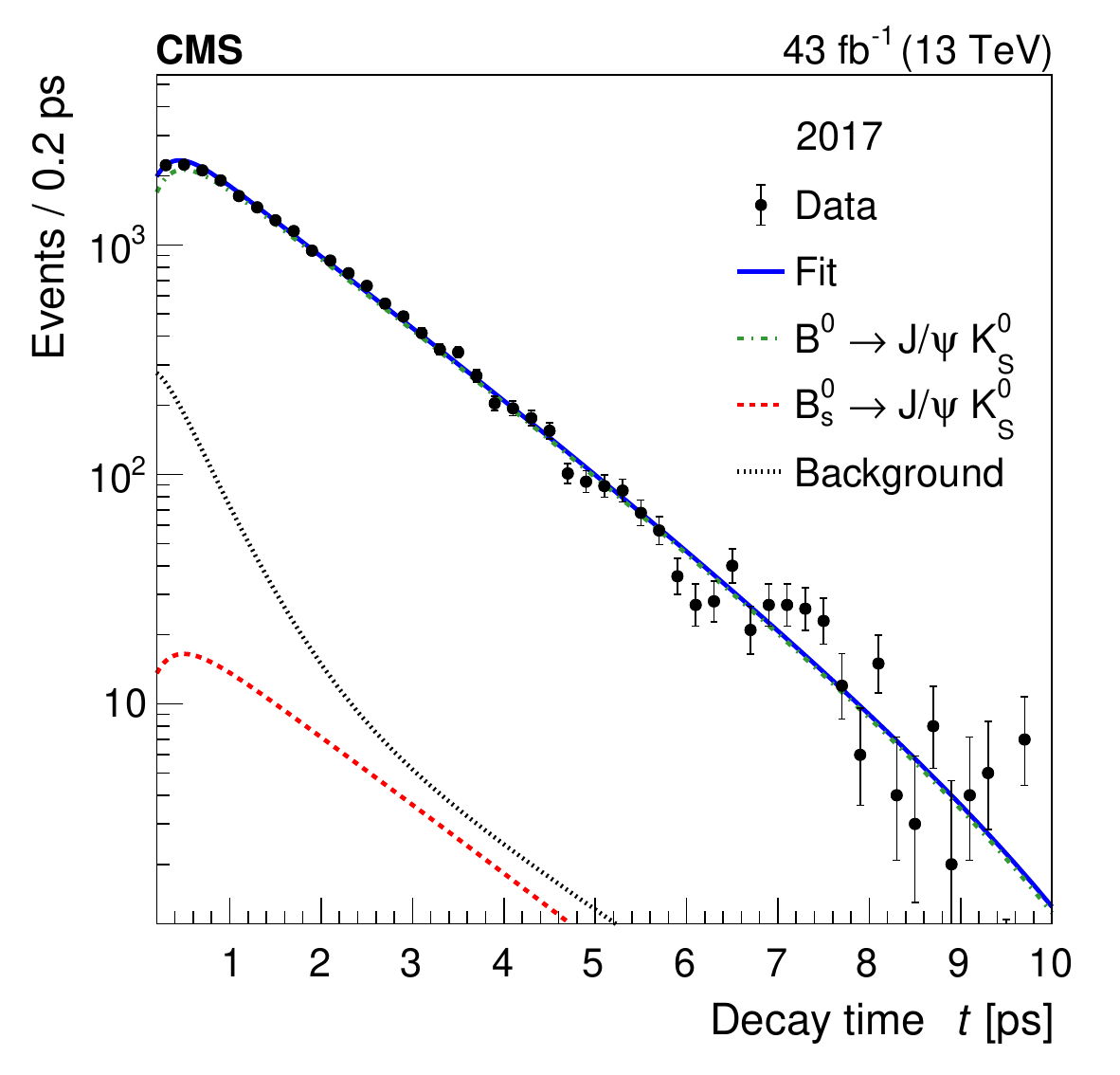}
    \includegraphics[width=0.32\textwidth]{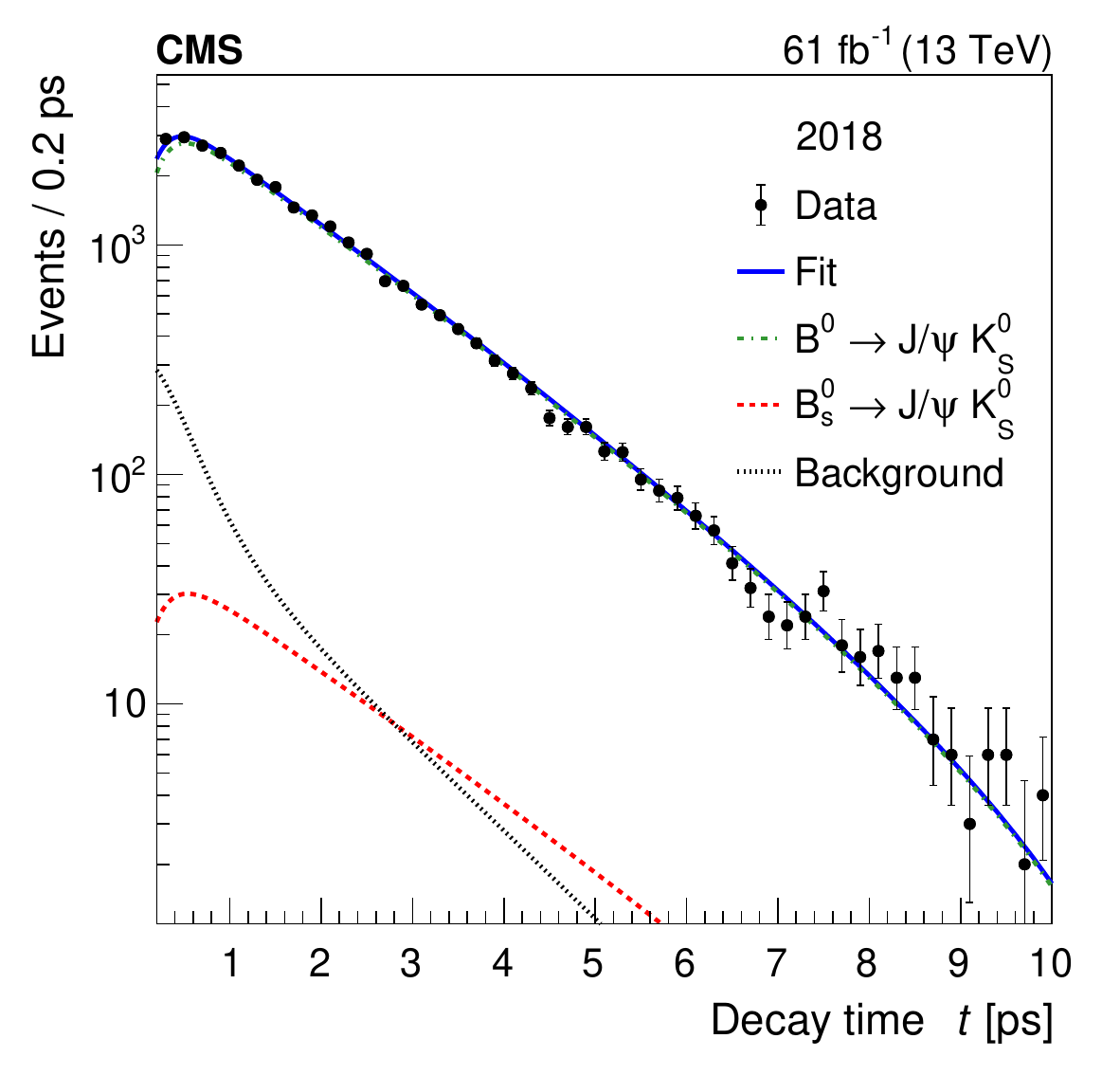}
    \caption{Distributions of the \jpsiks invariant mass (upper) and 
    decay time (lower) from data (points), along with the projections from the
    2D UML fit for each year of data taking. The vertical bars on the data points indicate the statistical uncertainty. The dashed, dotted-dashed, dotted, and solid lines represent the signal, 
    control channel, background, and total fit contributions, respectively.}
\label{fig:007}
\end{figure}

The invariant mass and decay time projection plots from the 2D UML fit are
presented in Fig.~\ref{fig:007}. The three plots in the upper row correspond to invariant 
mass projections for the 2016, 2017, and 2018 data-taking periods, respectively. 
The lower row displays the decay time projections for the data collected during the same periods.

\clearpage

\section{Subrange projection plots} \label{subrange_proj}
\begin{figure}[!h]
  \centering
    \includegraphics[width=0.32\textwidth]{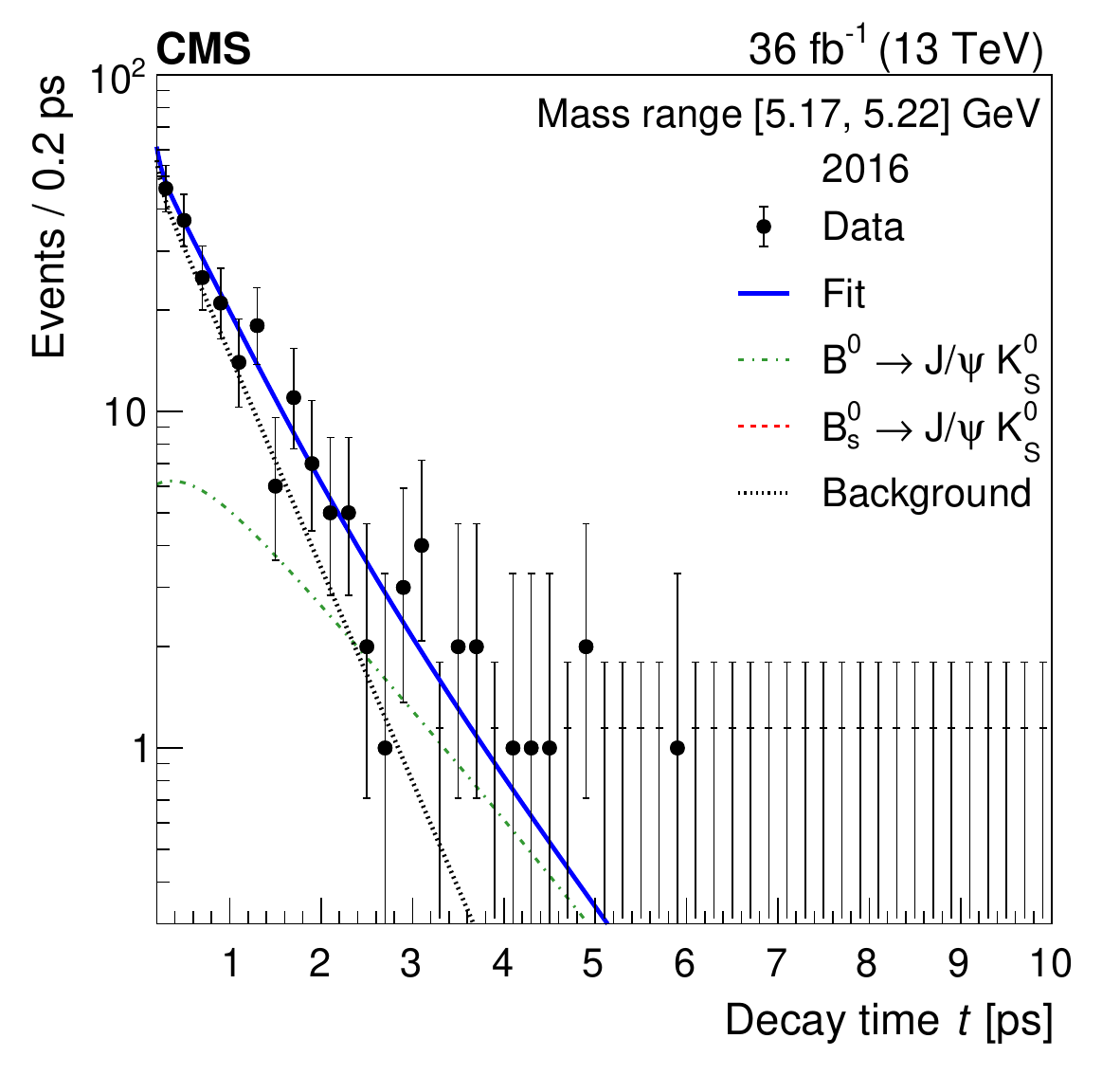}
    \includegraphics[width=0.32\textwidth]{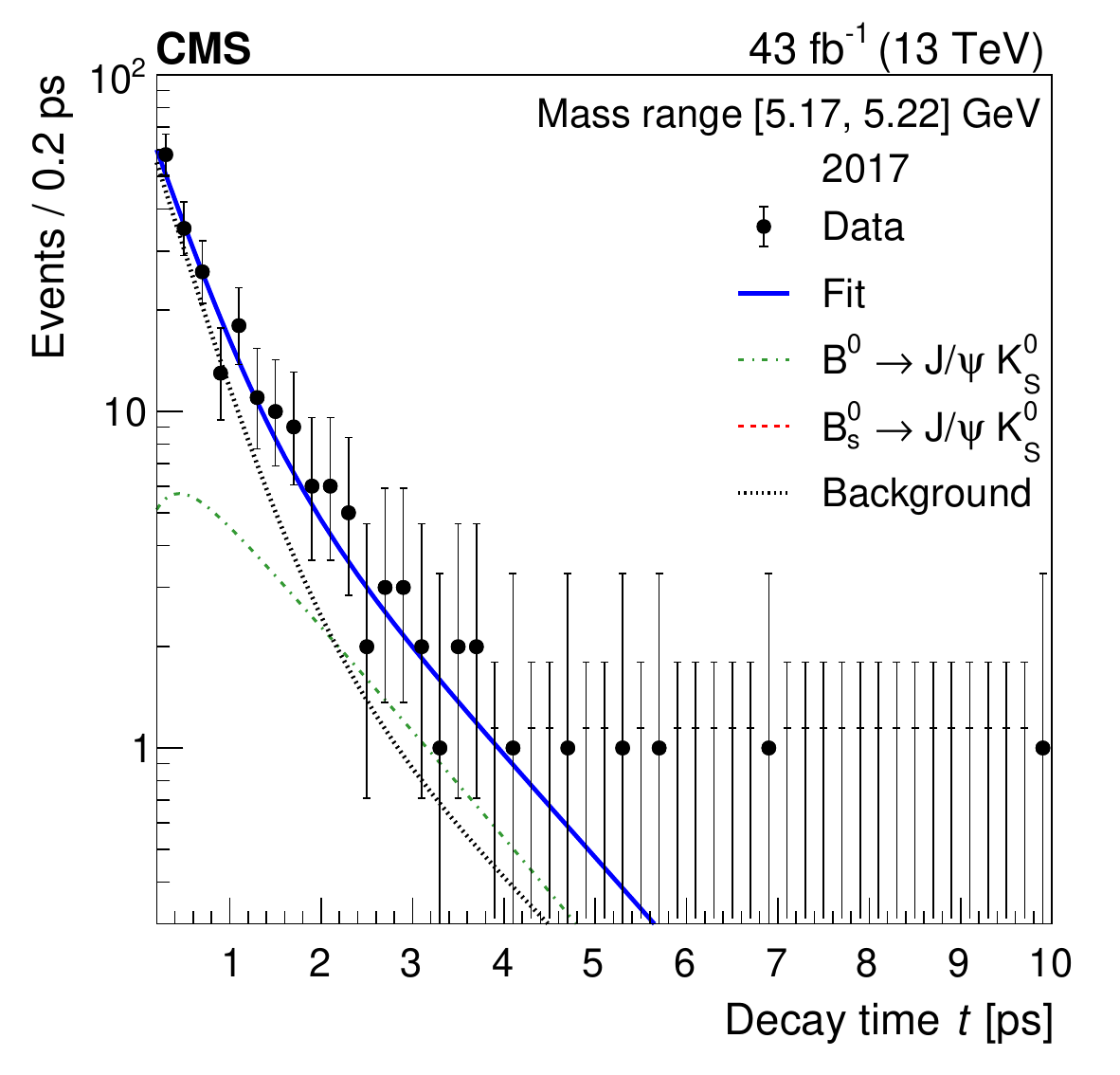}
    \includegraphics[width=0.32\textwidth]{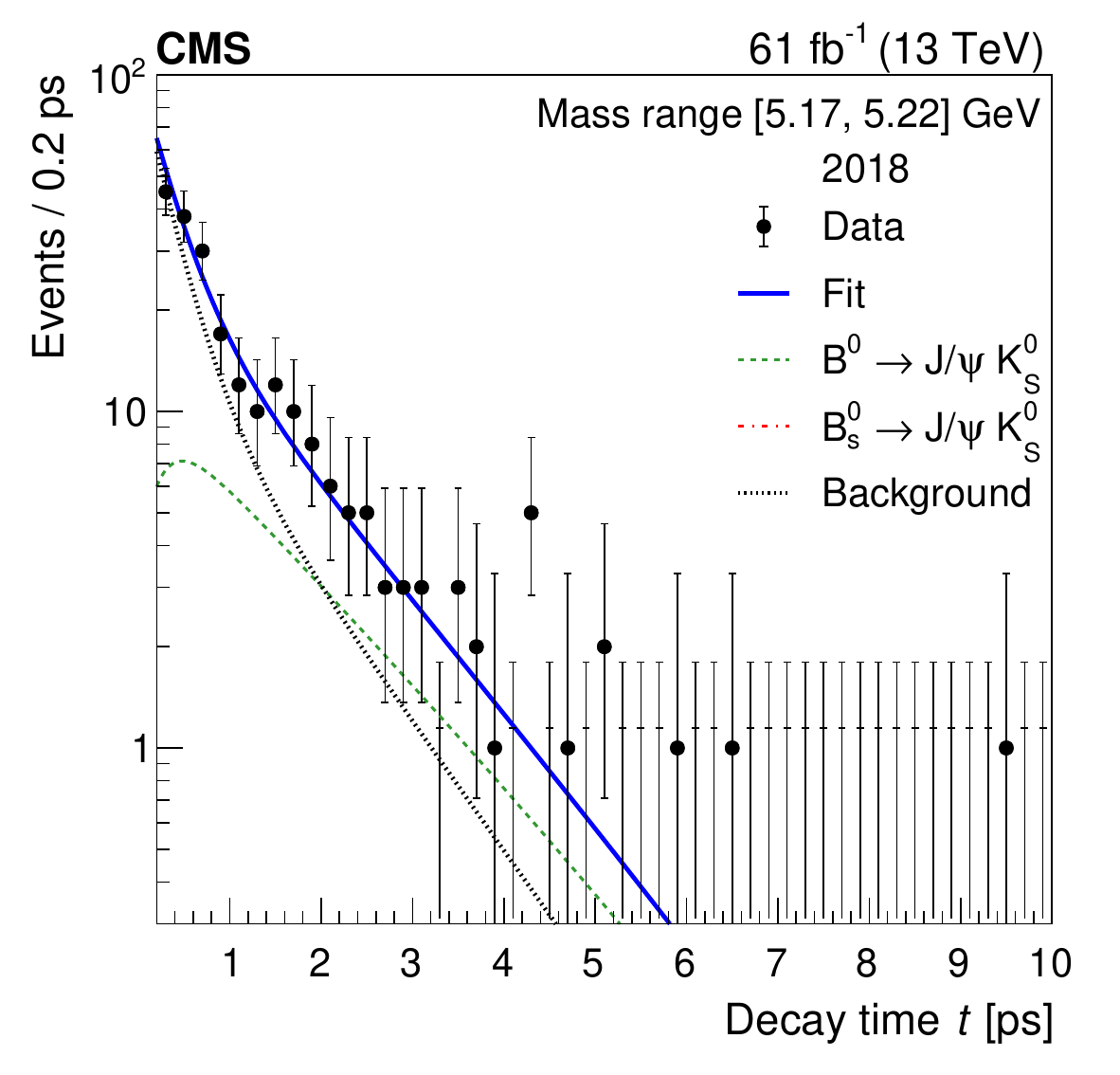}
    \includegraphics[width=0.32\textwidth]{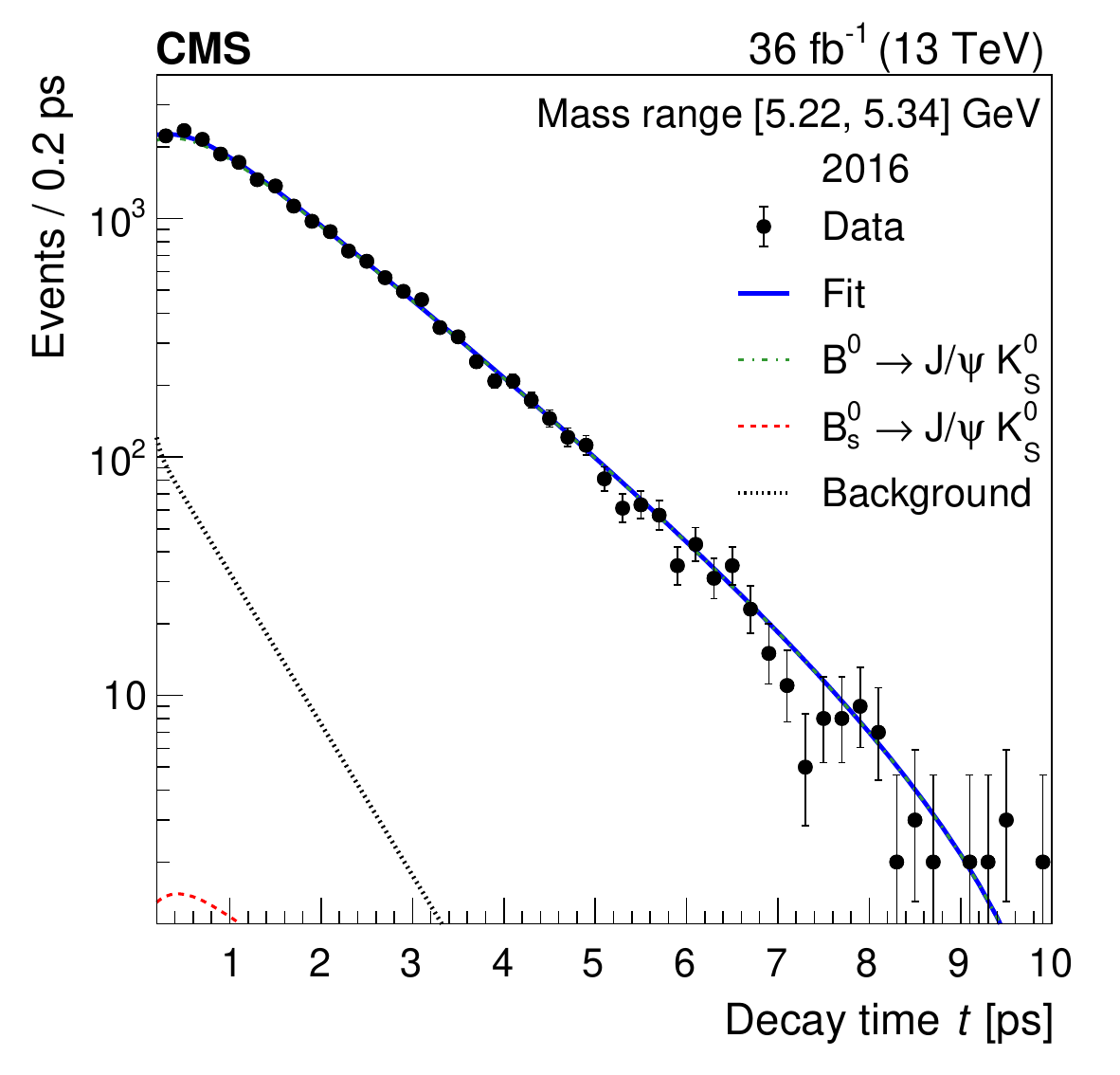}
    \includegraphics[width=0.32\textwidth]{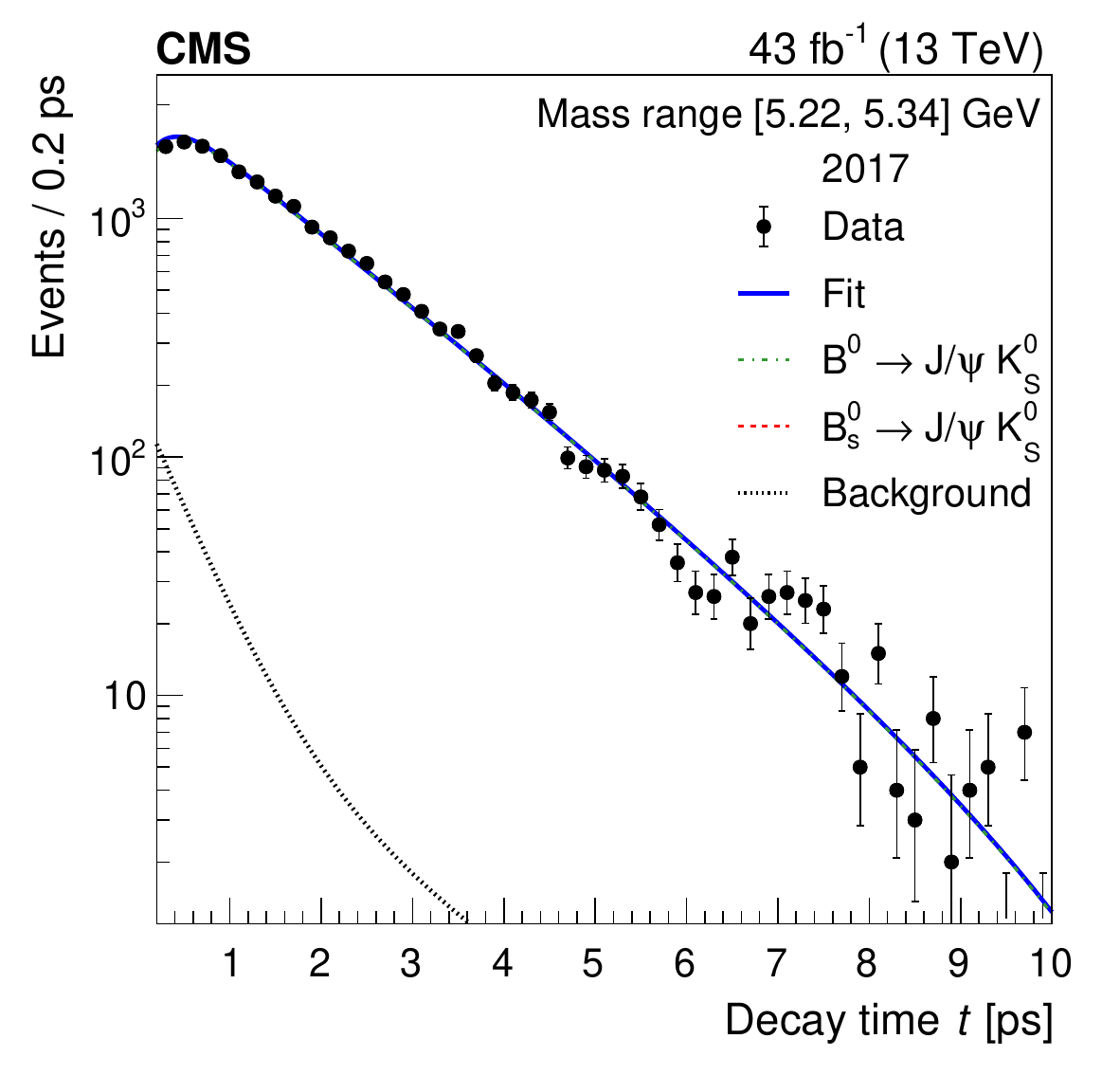}
    \includegraphics[width=0.32\textwidth]{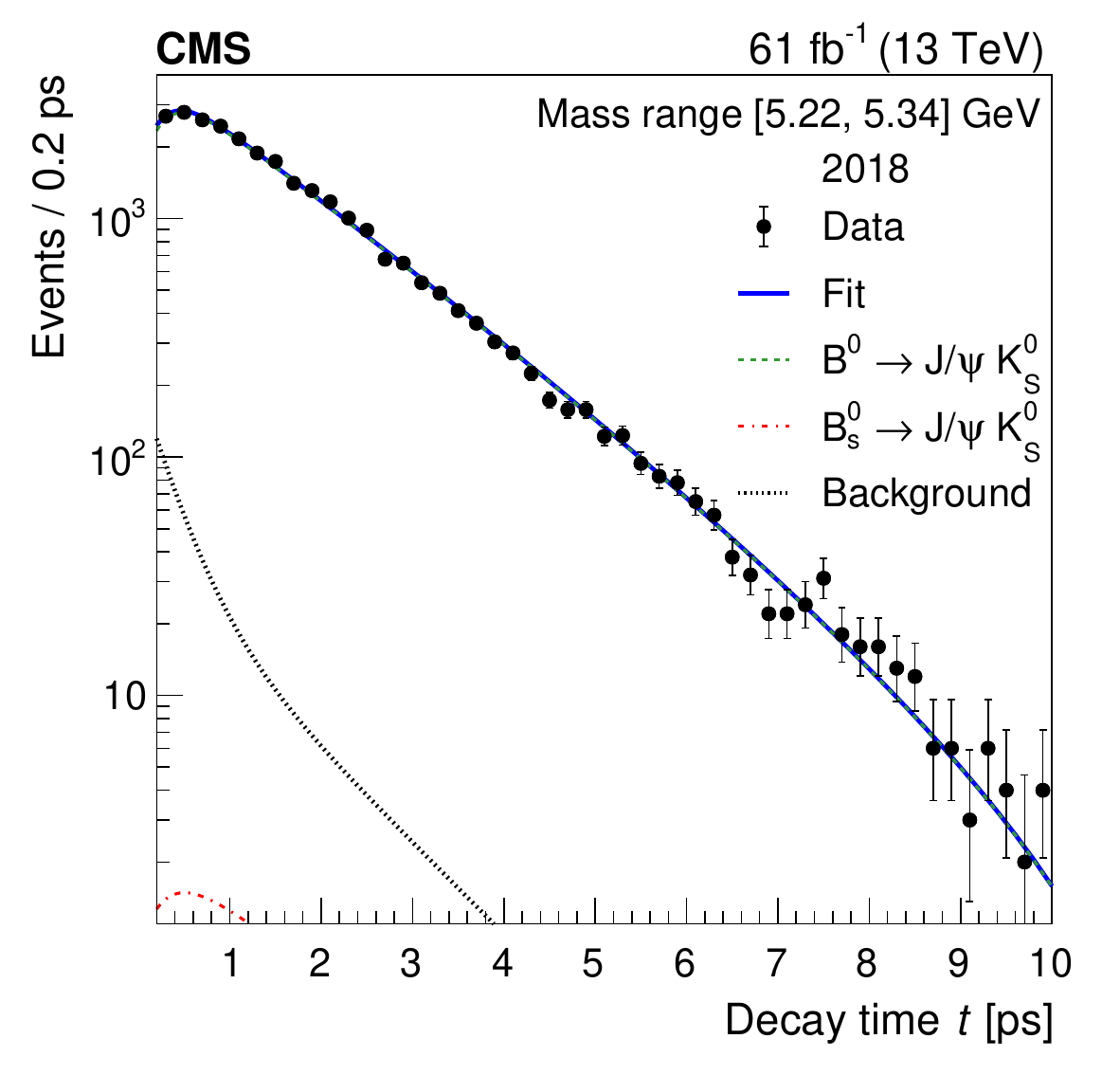}
    \includegraphics[width=0.32\textwidth]{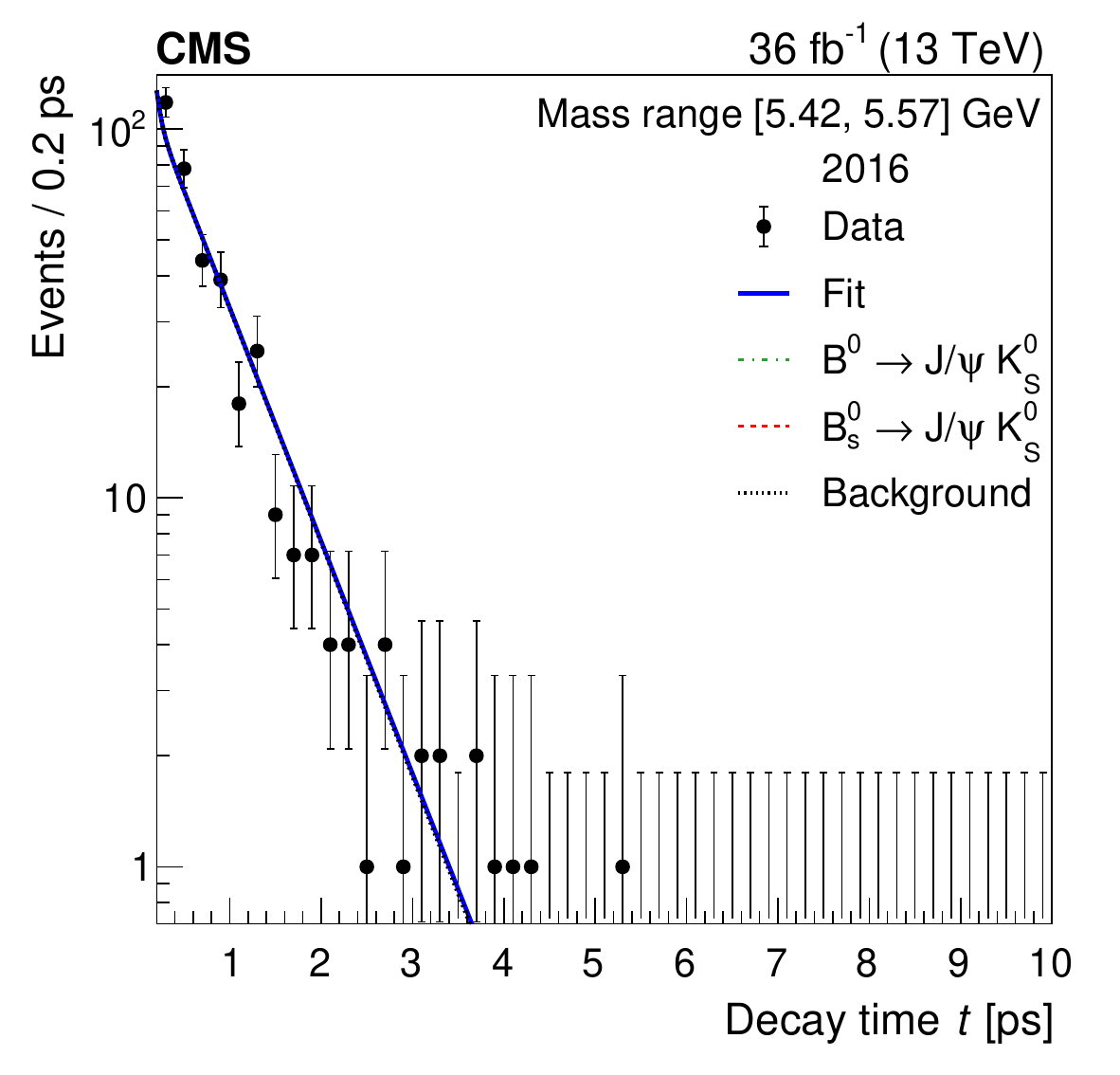}
    \includegraphics[width=0.32\textwidth]{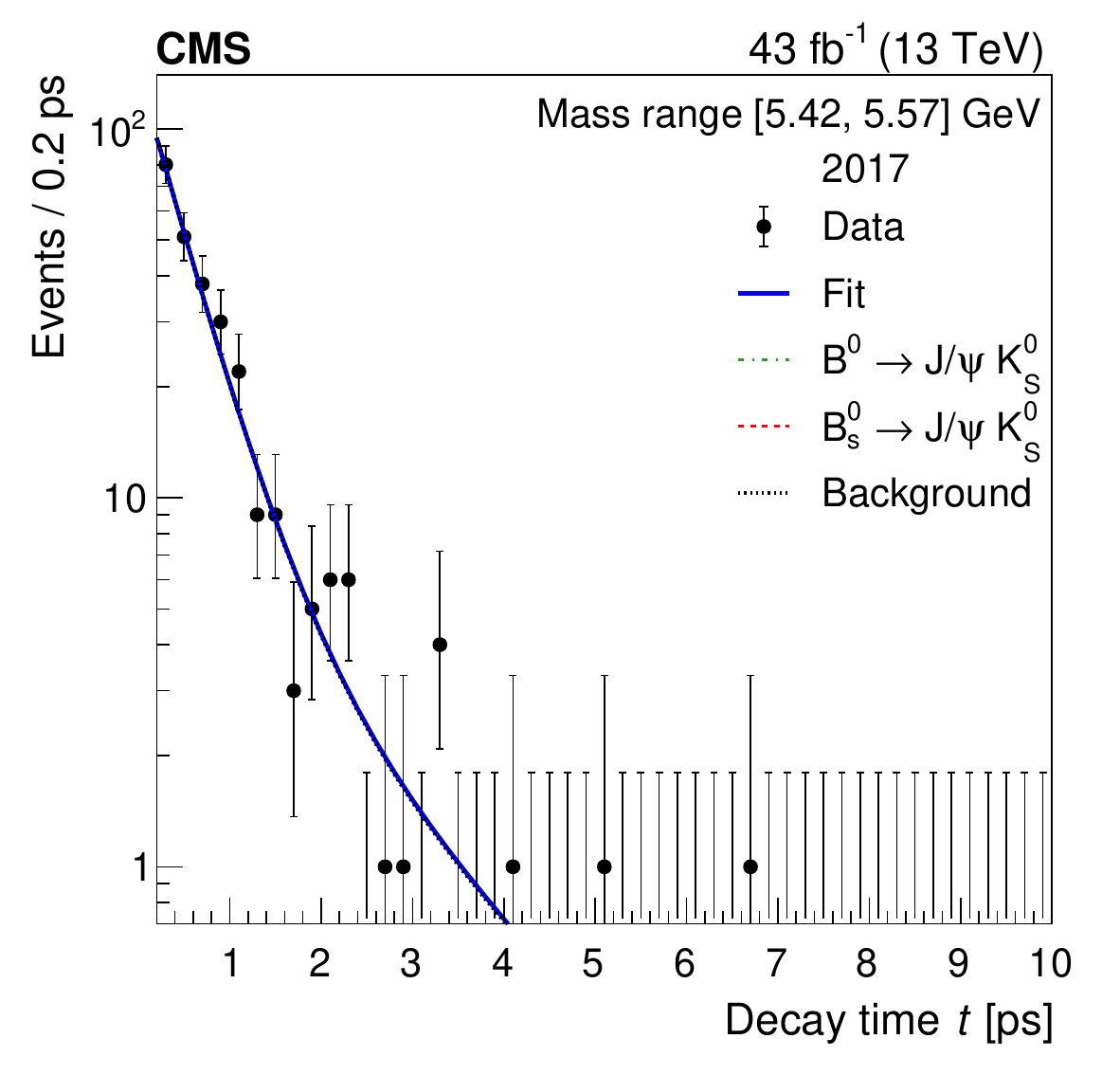}
    \includegraphics[width=0.32\textwidth]{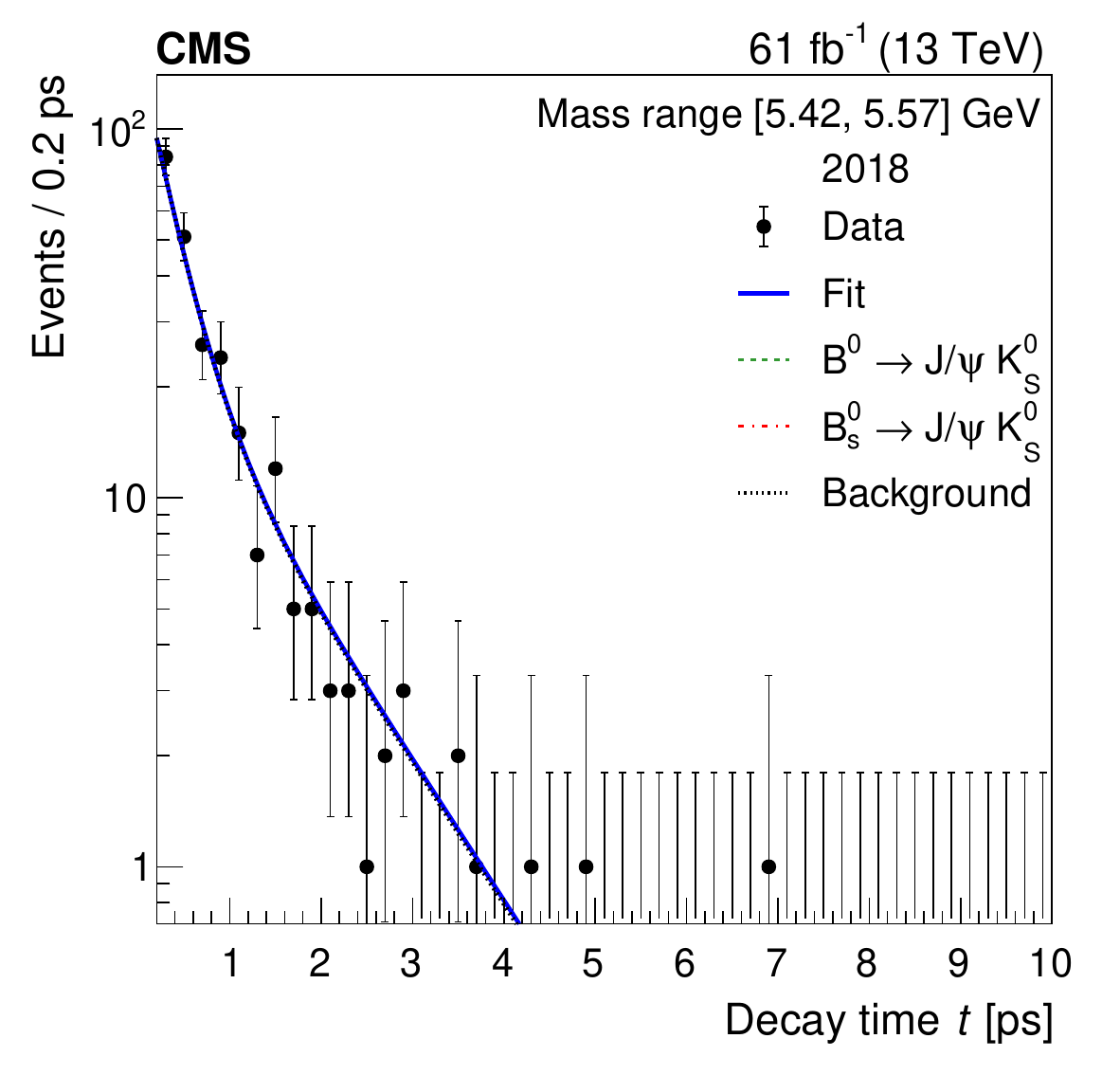}
\caption{The decay time distribution for events with \jpsiks
invariant mass in the range $5.17 < m < 5.22$ (upper), $5.22 < m < 5.34$ (center), and $5.42 < m < 5.57$
(lower)\GeV and the fit results for the 
2016 (left), 2017 (middle), and 2018 (right) data-taking years.}
\label{fig:rangesM}
\end{figure}

\begin{figure}[htb]
  \centering
    \includegraphics[width=0.32\textwidth]{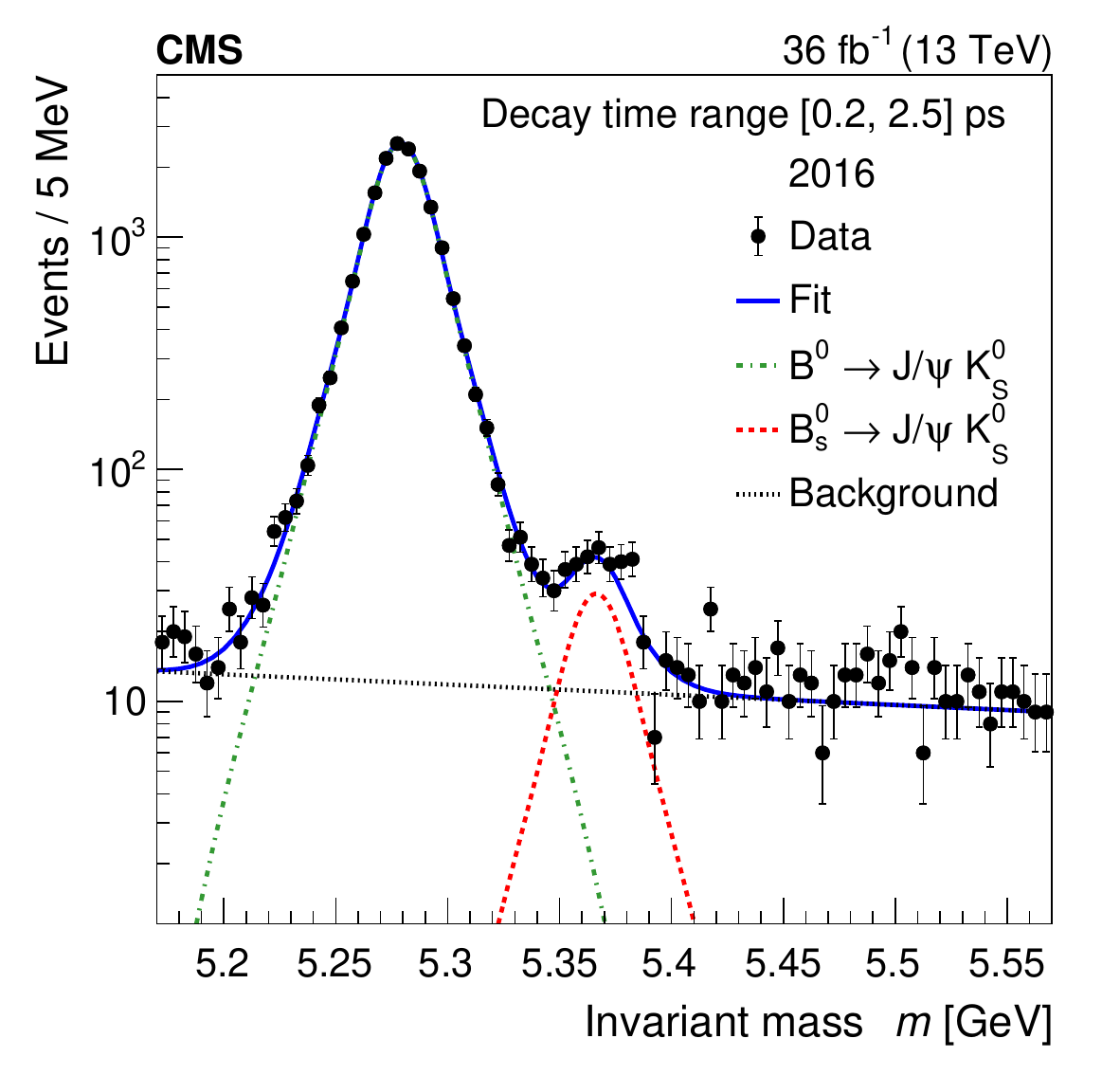}
    \includegraphics[width=0.32\textwidth]{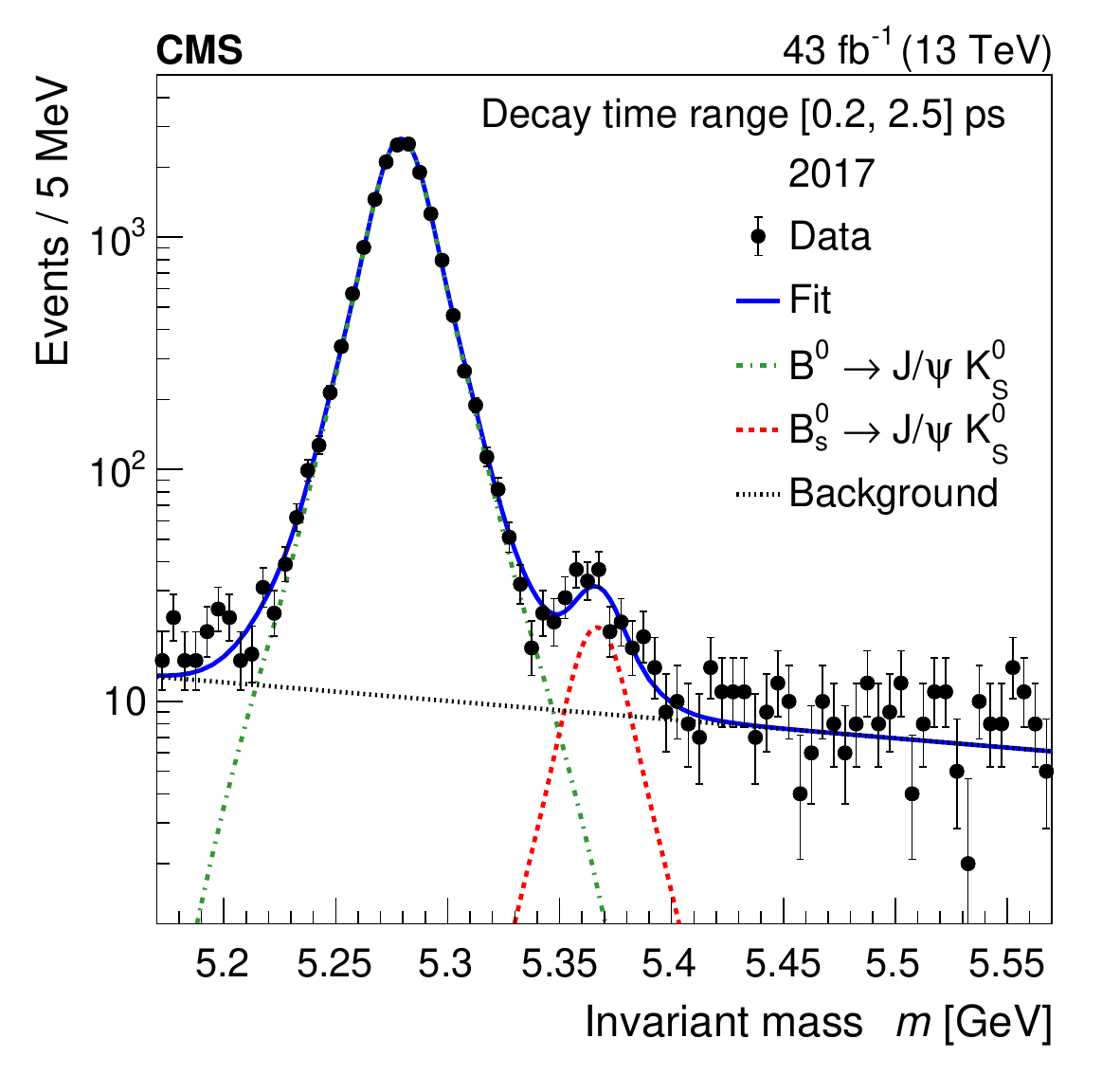}
    \includegraphics[width=0.32\textwidth]{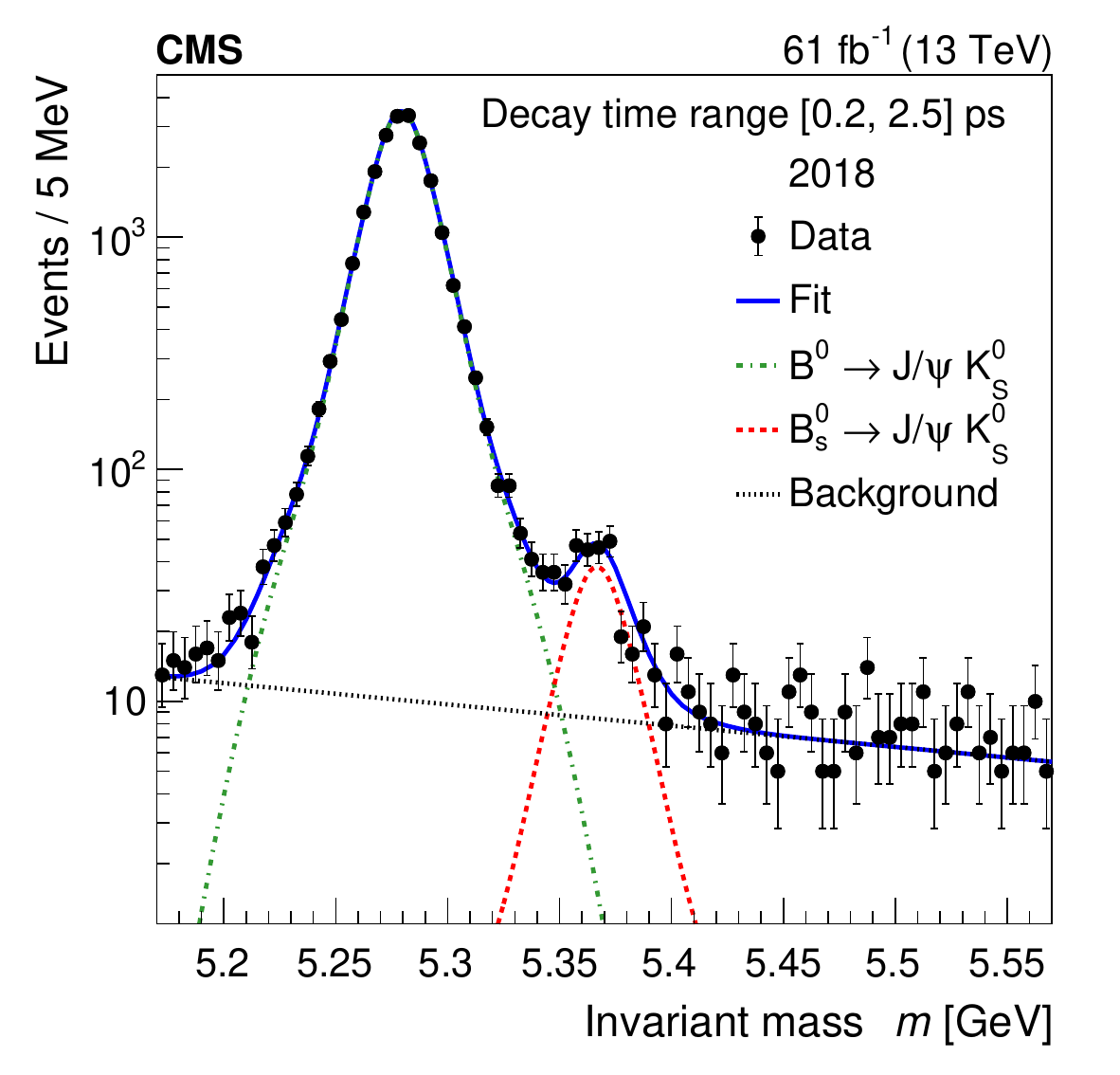}
    \includegraphics[width=0.32\textwidth]{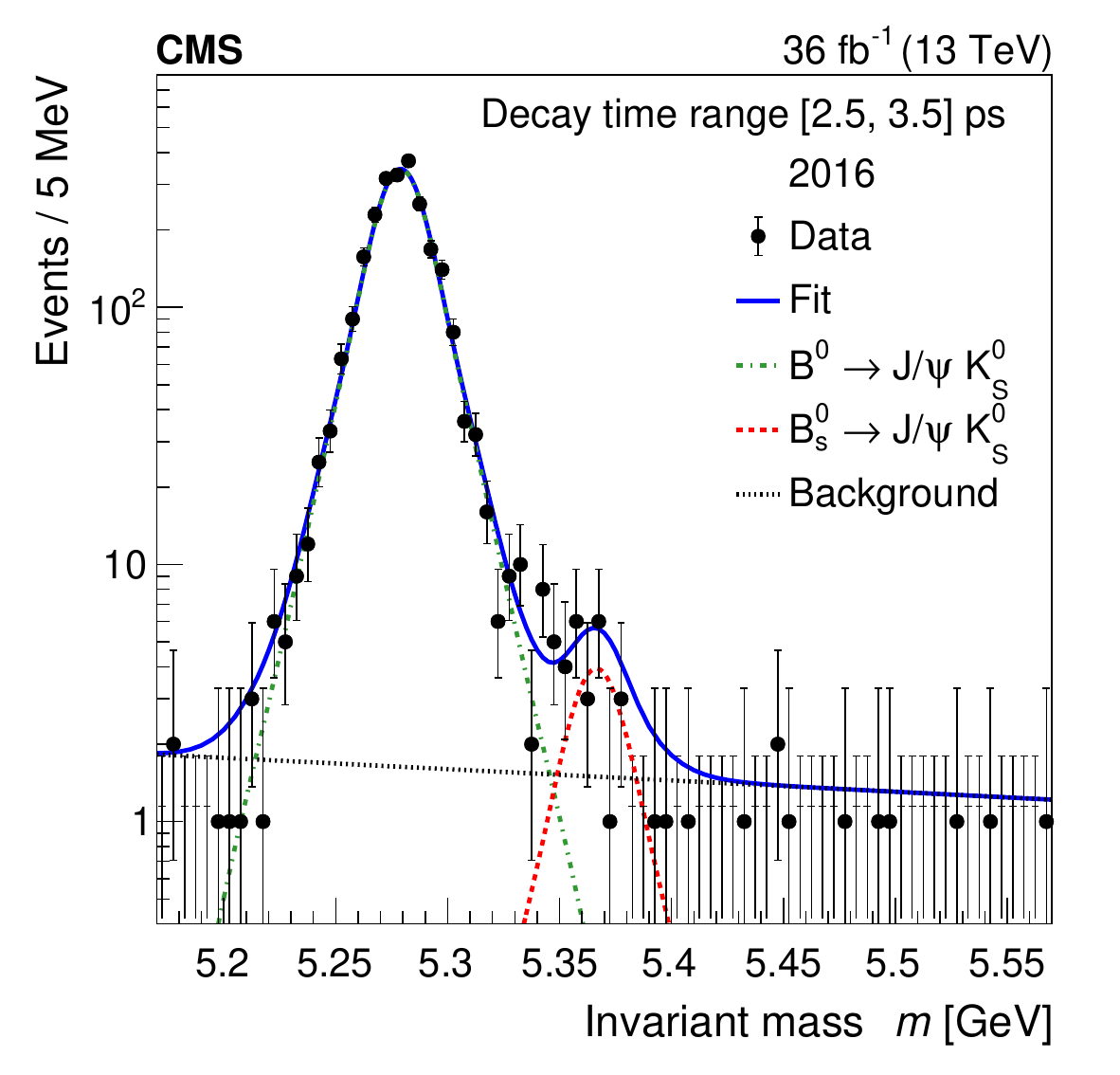}
    \includegraphics[width=0.32\textwidth]{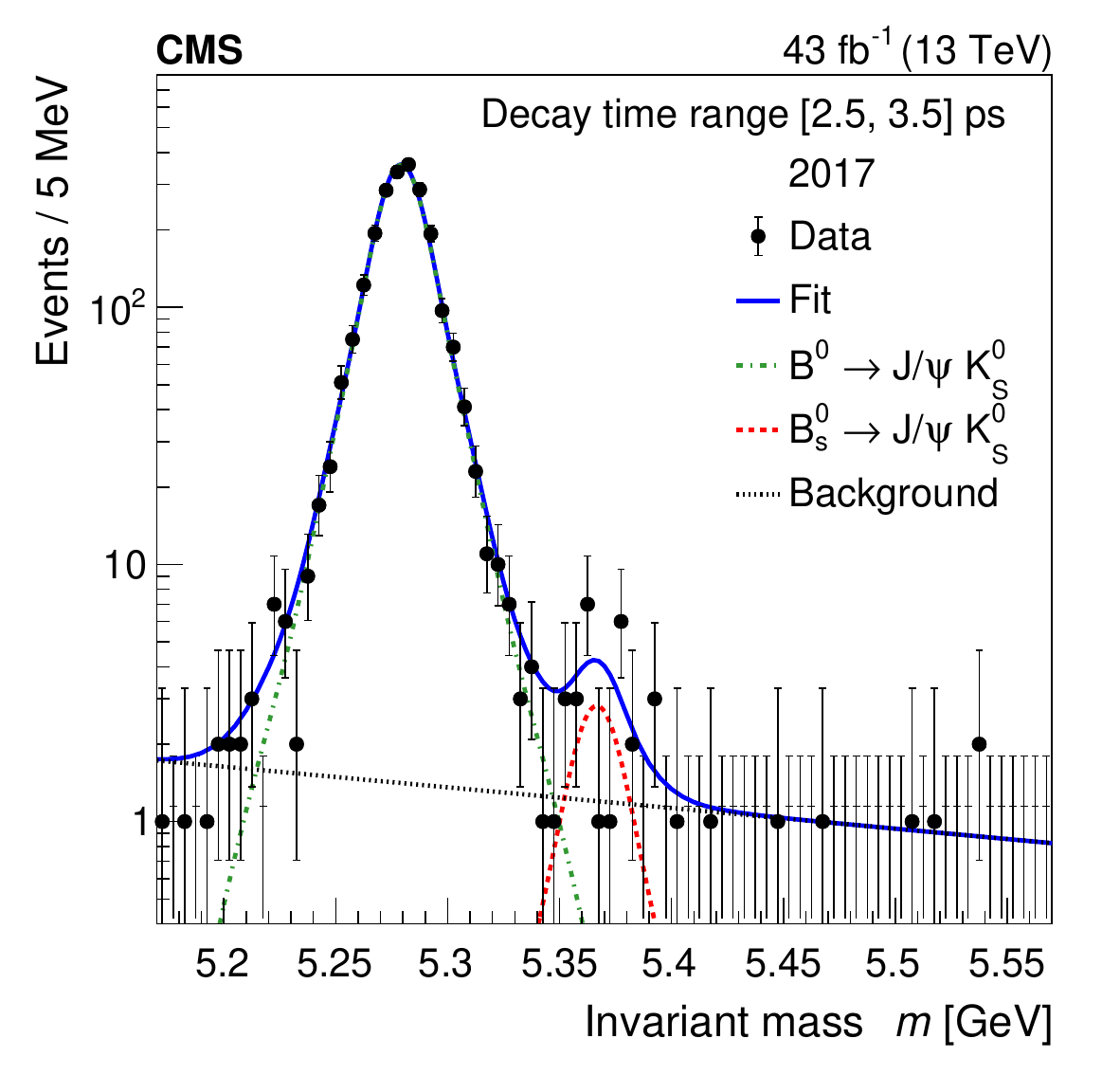}
    \includegraphics[width=0.32\textwidth]{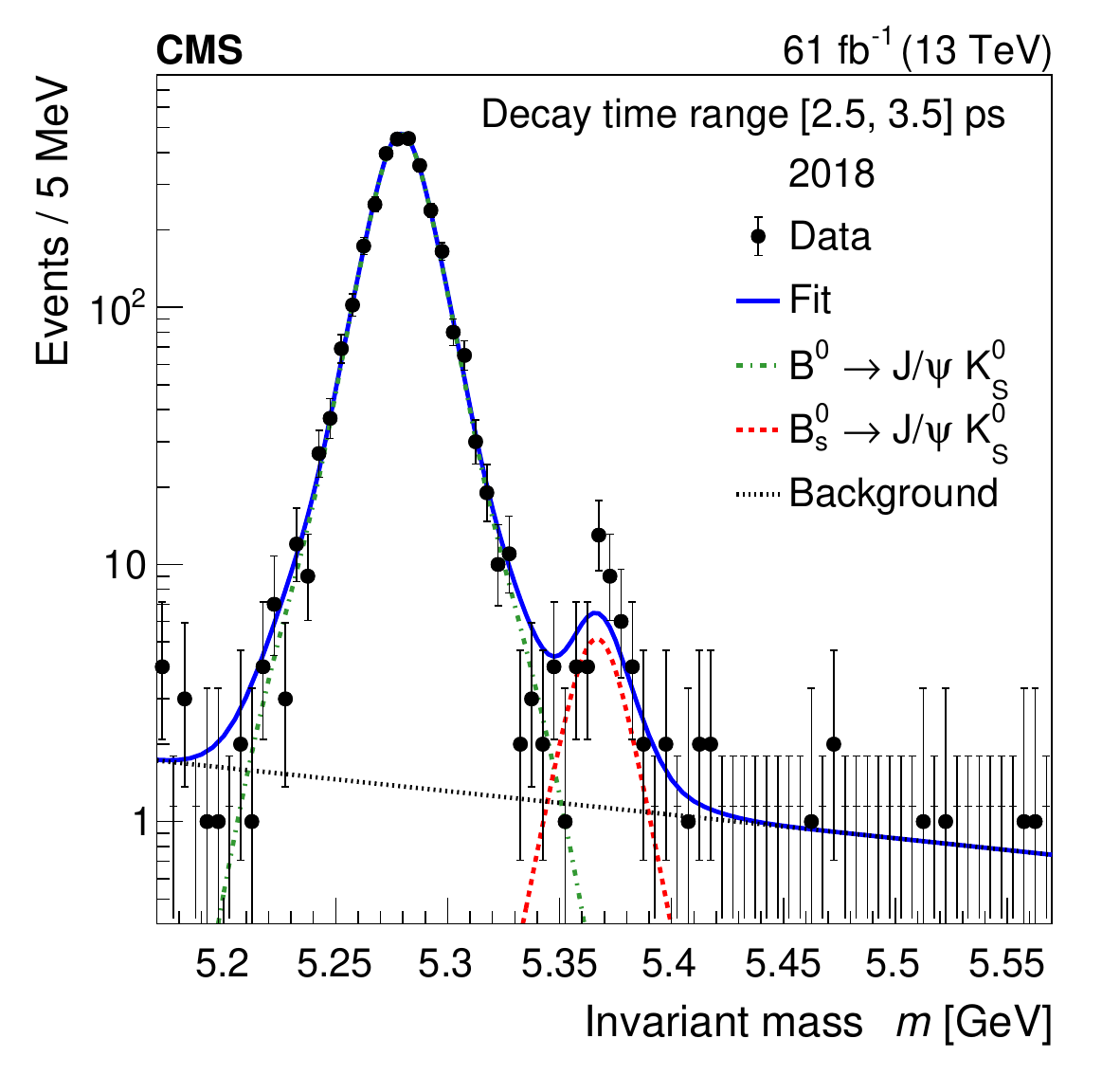}
    \includegraphics[width=0.32\textwidth]{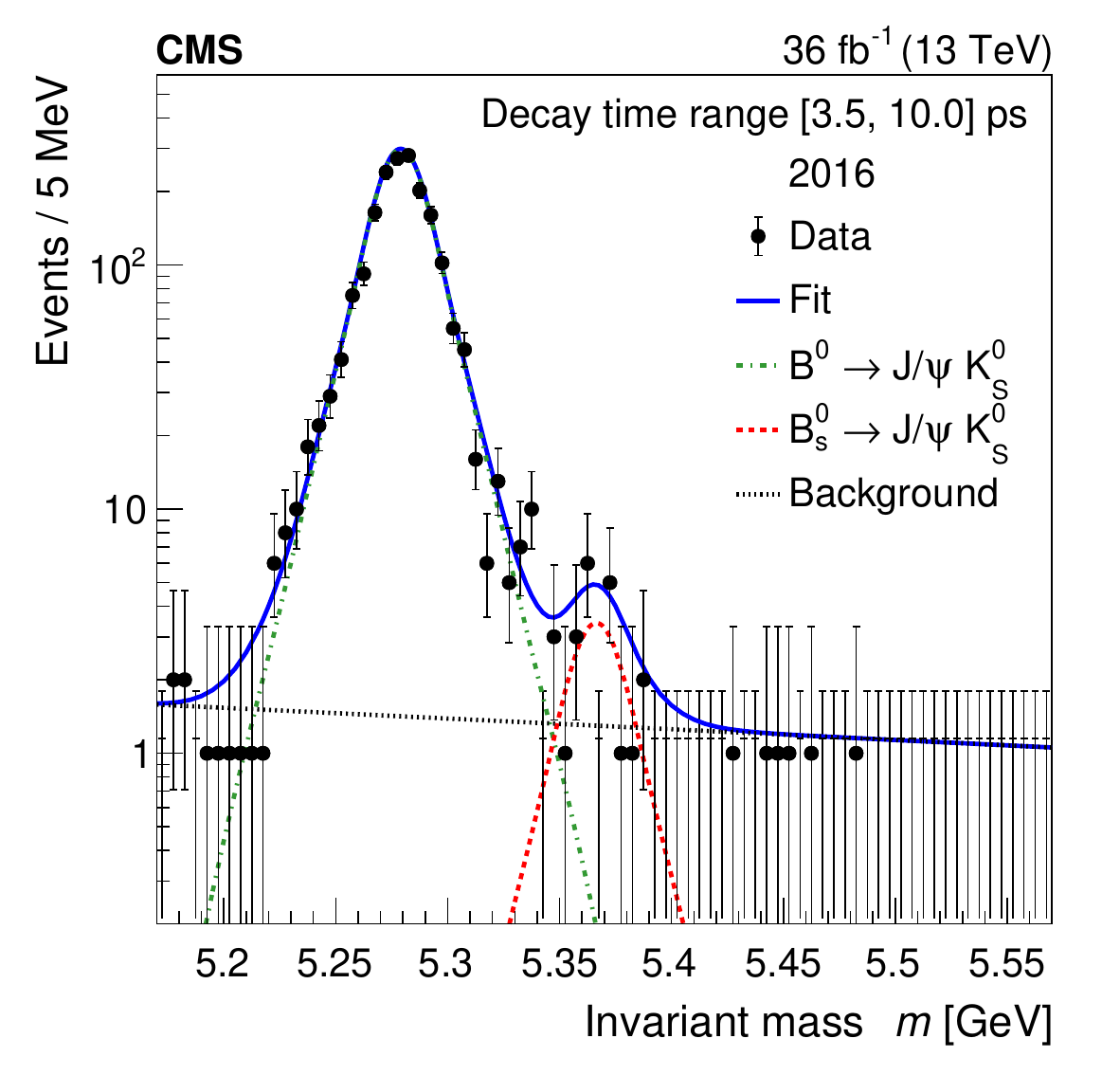}
    \includegraphics[width=0.32\textwidth]{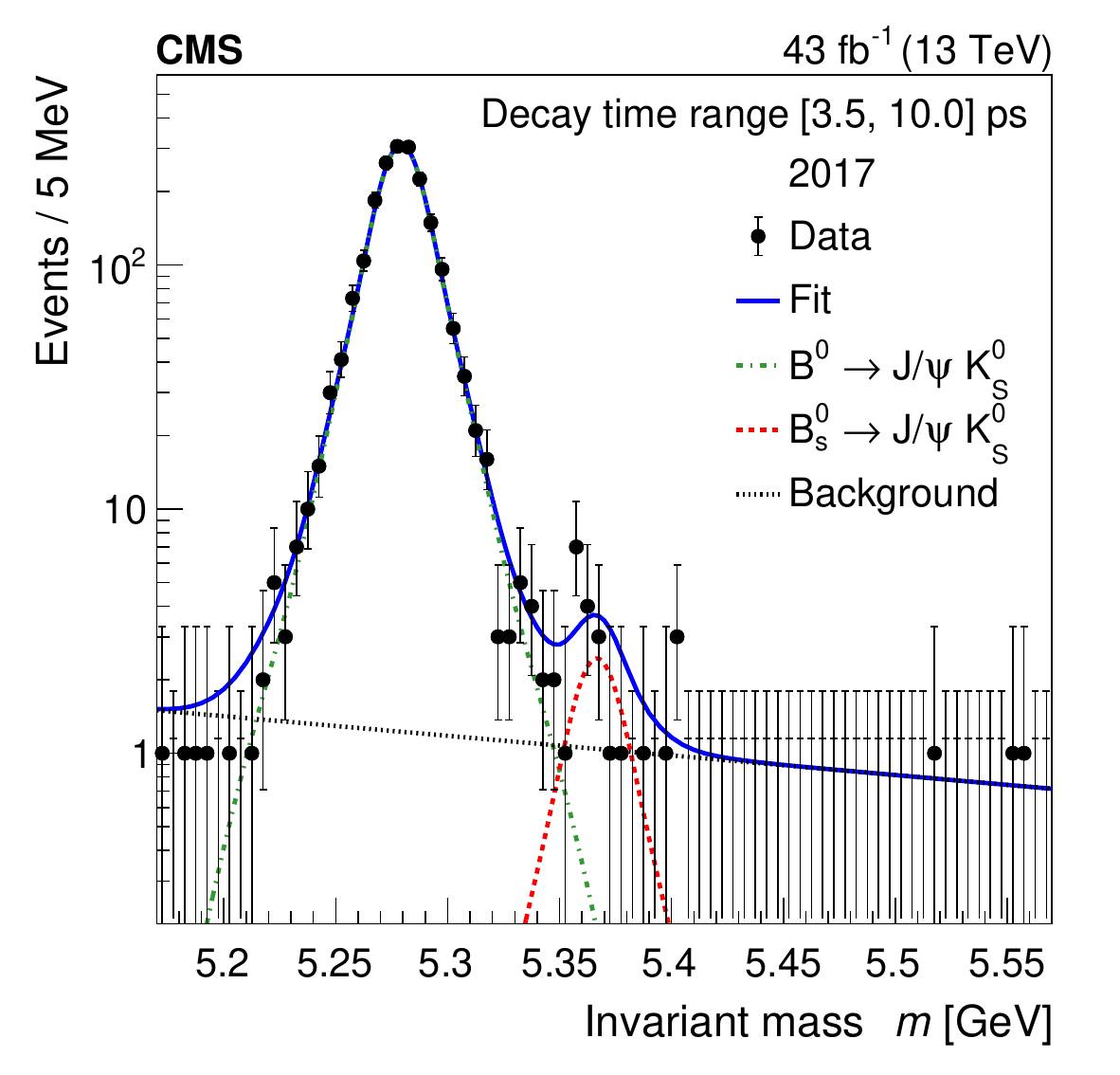}
    \includegraphics[width=0.32\textwidth]{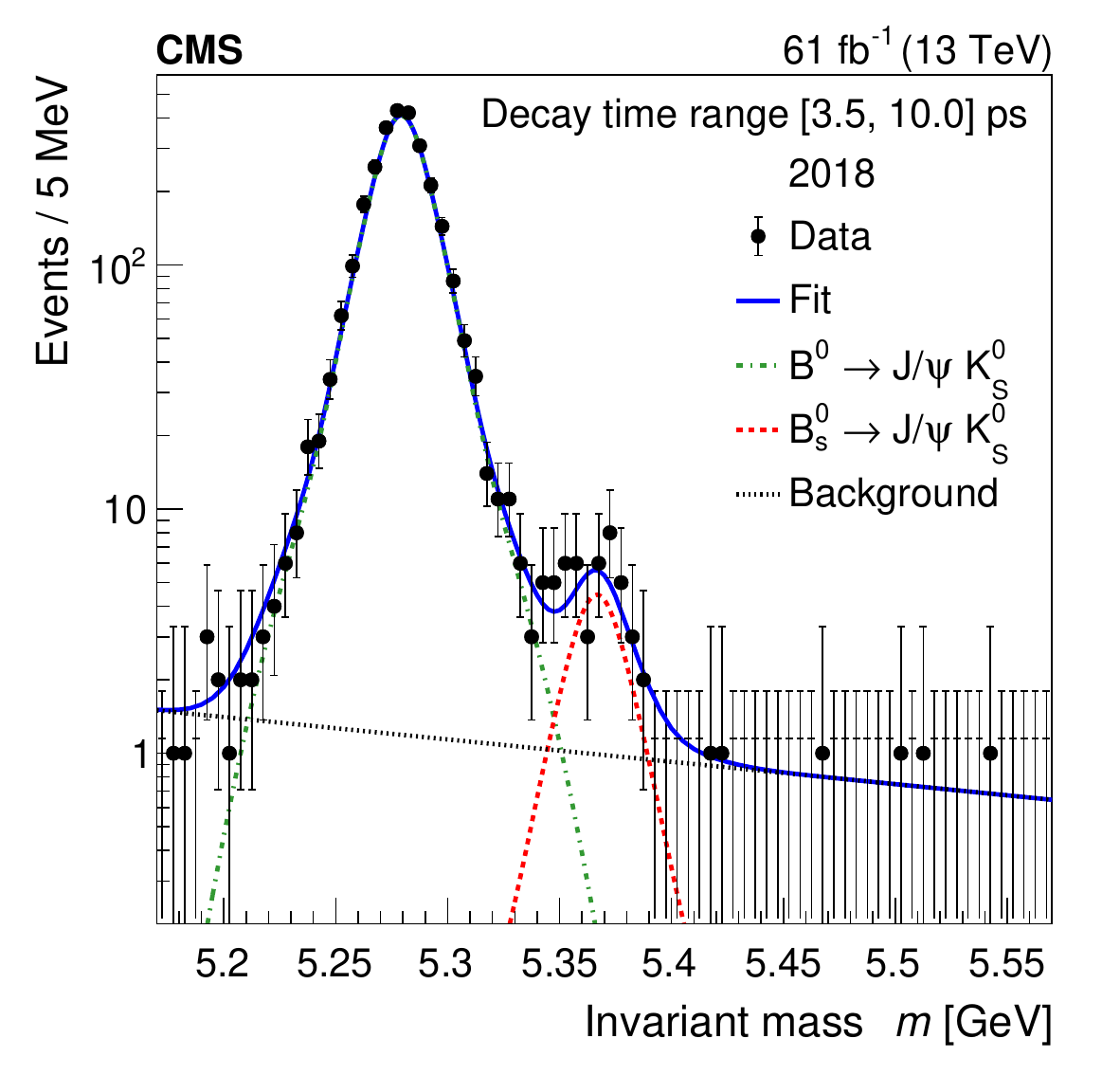}
\caption{The invariant mass distribution for events with \jpsiks
decay time in the range $0.2 < t < 2.5$ (upper), $2.5 < t < 3.5$ (center), and $3.5 < t < 10.0$
(lower)\ps and the fit results for the 2016 (left), 2017 (middle), and 2018 (right) data-taking years.}
\label{fig:rangesdt}
\end{figure}

Projections of the decay time distributions for three different \jpsiks invariant mass ranges 
and the results of the 2D UML fits
are presented in Fig.~\ref{fig:rangesM} for each of the three data-taking years. The
corresponding invariant mass projections for different decay time
ranges are shown in Fig.~\ref{fig:rangesdt}.

\cleardoublepage \section{The CMS Collaboration \label{app:collab}}\begin{sloppypar}\hyphenpenalty=5000\widowpenalty=500\clubpenalty=5000\input{BPH-22-001-public-authorlist.tex}\end{sloppypar}
\end{document}

%% file: BPH-22-001-public-authorlist.tex
\cmsinstitute{Yerevan Physics Institute, Yerevan, Armenia}
{\tolerance=6000
A.~Hayrapetyan, A.~Tumasyan\cmsAuthorMark{1}\cmsorcid{0009-0000-0684-6742}
\par}
\cmsinstitute{Institut f\"{u}r Hochenergiephysik, Vienna, Austria}
{\tolerance=6000
W.~Adam\cmsorcid{0000-0001-9099-4341}, J.W.~Andrejkovic, T.~Bergauer\cmsorcid{0000-0002-5786-0293}, S.~Chatterjee\cmsorcid{0000-0003-2660-0349}, K.~Damanakis\cmsorcid{0000-0001-5389-2872}, M.~Dragicevic\cmsorcid{0000-0003-1967-6783}, P.S.~Hussain\cmsorcid{0000-0002-4825-5278}, M.~Jeitler\cmsAuthorMark{2}\cmsorcid{0000-0002-5141-9560}, N.~Krammer\cmsorcid{0000-0002-0548-0985}, A.~Li\cmsorcid{0000-0002-4547-116X}, D.~Liko\cmsorcid{0000-0002-3380-473X}, I.~Mikulec\cmsorcid{0000-0003-0385-2746}, J.~Schieck\cmsAuthorMark{2}\cmsorcid{0000-0002-1058-8093}, R.~Sch\"{o}fbeck\cmsorcid{0000-0002-2332-8784}, D.~Schwarz\cmsorcid{0000-0002-3821-7331}, M.~Sonawane\cmsorcid{0000-0003-0510-7010}, W.~Waltenberger\cmsorcid{0000-0002-6215-7228}, C.-E.~Wulz\cmsAuthorMark{2}\cmsorcid{0000-0001-9226-5812}
% ...
\par}
\vskip\cmsinstskip
\dag:~Deceased\\
$^{1}$Also at Yerevan State University, Yerevan, Armenia\\
$^{2}$Also at TU Wien, Vienna, Austria\\
% ...
